\documentclass[apj]{emulateapj}
\usepackage{apjfonts}

\newcommand\kms           {km~s$^{-1}$}

\newcommand{\tnm}         {\tablenotemark}
\newcommand{\tnt}         {\tablenotetext}
\newcommand{\hii}         {\ion{H}{2}}
\newcommand{\rarr}        {\ensuremath{\rightarrow}}

\begin{document}
\shorttitle{High Spectral Resolution OH Masers in W3(OH)}
\shortauthors{Fish, Brisken, \& Sjouwerman}
\title{A Very High Spectral Resolution Study of Ground-State OH Masers
 in W3(OH)}
\author{Vincent L.~Fish\altaffilmark{1},
        Walter F.~Brisken, \&
        Lor\'{a}nt O.~Sjouwerman}

\affil{National Radio Astronomy Observatory, P. O. Box O, 1003
Lopezville Road, Socorro, NM 87801}
\email{vfish@nrao.edu, wbrisken@nrao.edu, lsjouwer@nrao.edu}

\altaffiltext{1}{Jansky Fellow.}

\begin{abstract}
We present VLBA observations of the ground-state hydroxyl masers in
W3(OH) at 0.02~\kms\ spectral resolution.  Over 250 masers are
detected, including 56 Zeeman pairs.  Lineshapes are predominantly
Gaussian or combinations of several Gaussians, with normalized
deviations typically of the same magnitude as in masers in other
species.  Typical FWHM maser linewidths are 0.15 to 0.38~\kms\ and are
larger in the 1665 MHz transition than in the other three ground-state
transitions.  The satellite-line 1612 and 1720 MHz masers show no
evidence of $\sigma^{\pm2,3}$ components.  The spatial positions of
most masers are seen to vary across the line profile, with many spots
showing clear, organized positional gradients.  Equivalent
line-of-sight velocity gradients in the plane of the sky typically
range from 0.01 to 1 \kms~AU$^{-1}$ (i.e., positional gradients of 1
to 100 AU~(\kms)$^{-1}$).  Small velocity gradients in the 1667 MHz
transition support theoretical predictions that 1667 MHz masers appear
in regions with small velocity shifts along the amplification length.
Deconvolved maser spot sizes appear to be larger in the line wings but
do not support a spherical maser geometry.
\end{abstract}


\section{Introduction}
\label{introduction}

\begin{figure}
\begin{center}
\resizebox{\hsize}{!}{\includegraphics{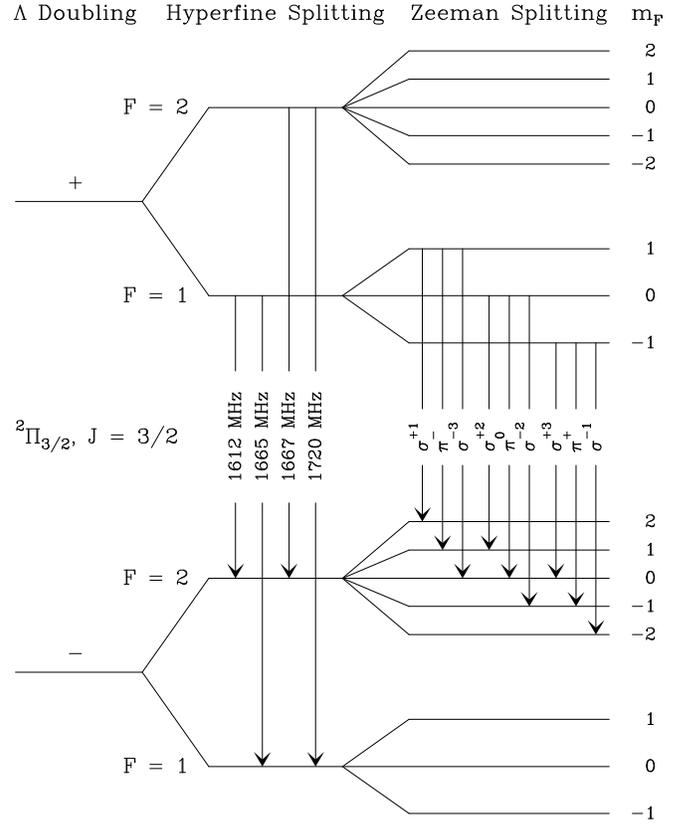}}
\end{center}
\caption{Satellite-line Zeeman components for the 1612 MHz transition
  of OH, adapted from \citet{hansen82}.  Because the Zeeman splitting
  is different for different $F$ levels, the $\sigma$ and $\pi$
  components for an $F$-nonconserving transition are nondegenerate.
  In the presence of a magnetic field, a 1612 MHz line will split into
  six $\sigma$ components and three $\pi$ components.  In the presence
  of a magnetic field along the line of sight, the $\sigma^+$
  components are left circularly polarized and the $\sigma^-$
  components are right circularly polarized.  The Zeeman splitting for
  the 1720 MHz transition is analogous.  Splittings are not to
  scale.\label{fig-split}}
\end{figure}

\begin{figure}
\begin{center}
\resizebox{\hsize}{!}{\includegraphics{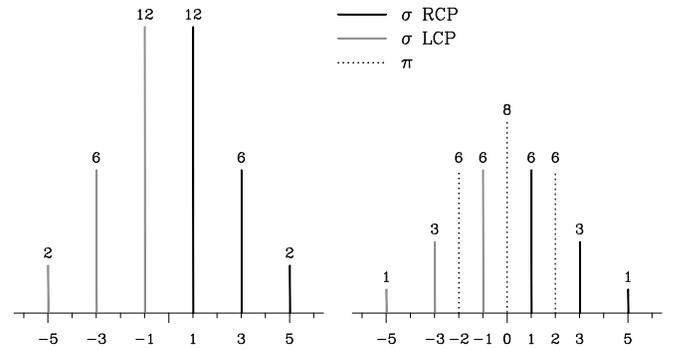}}
\end{center}
\caption{Zeeman splitting pattern of 1612 and 1720 MHz masers in LTE,
  adapted from \citet{davies74}.  \emph{Left}: Splitting pattern when
  the magnetic field is along the line of sight.  No $\pi$ components
  are present in the spectrum.  Numbers above the lines indicate the
  relative intensities.  Numbers along the axis indicate the velocity
  shift of the Zeeman components from the systemic velocity, in units
  of 0.061 \kms~mG$^{-1}$ for the 1612 MHz transition and 0.057
  \kms~mG$^{-1}$ for the 1720 MHz transition.  In the presence of a
  positive magnetic field (i.e., oriented in the hemisphere pointing
  away from the observer), the RCP $\sigma^-$ components are shifted
  to higher velocity than the LCP $\sigma^+$ components; a negative
  magnetic field produces the reverse pattern.  \emph{Right}:
  Splitting pattern when the magnetic field is nearly in the plane of
  the sky.  Intensities of the $\sigma$ components are in the same
  ratio, though a factor of two weaker than for the case of the
  magnetic field along the line of sight.  Three $\pi$ components also
  appear.
  \label{fig-intensities}}
\end{figure}

In the presence of a magnetic field, the degeneracy of magnetic
sublevels of a molecule is broken due to the Zeeman
effect.  Zeeman splitting of the hydroxyl radical (OH) is often used
to infer magnetic field strengths, both in masers
\citep[e.g.,][]{davies66} and in thermal gas
\citep[e.g.,][]{turner70}.  For the main-line, $F$-conserving
transitions, the line splits into one $\pi$ component at the systemic
velocity and two $\sigma$ components ($\sigma^+$ and $\sigma^-$)
shifted in opposite senses with respect to the systemic velocity.  For
transitions in which $\Delta F = \pm 1$, such as the 1612 MHz ($F = 1
\rarr 2$) and 1720 MHz ($F = 2 \rarr 1$) transitions of OH, the
splitting is more complicated (Figure \ref{fig-split}).  These
ground-state satellite lines split into six $\sigma$ components
($\sigma^{\pm 1,2,3}$) and three $\pi$ components ($\pi^0, \pi^\pm$),
with component intensities in local thermodynamic equilibrium (LTE)
being strongest for the innermost $\sigma^{\pm 1}$ components
(Figure \ref{fig-intensities}).  Excited-state satellite lines split
into a larger number of components; for instance, the 6016 and 6049
MHz lines each split into 15 different lines in the presence of a
magnetic field \citep{davies74}.

With the exception of a single marginal Zeeman triplet at the
$F$-conserving 1665 MHz transition in W75~N
\citep{hutawarakorn02,fish06}, a full Zeeman pattern has never been
observed in interstellar OH masers.  In most sources, no clear $\pi$
components are seen at all.  In the $F$-nonconserving satellite lines,
theoretical considerations of cross-relaxation among magnetic
sublevels due to trapped infrared radiation predict that even the
$\sigma^{\pm 2}$ and $\sigma^{\pm 3}$ components should not be
observable (\citealt{goldreich73b} as well as the discussion in
\citealt{lo75}).  Single-dish observations of the 1612 MHz OH
transition in Orion A are suggestive of the presence of $\sigma^{\pm
2}$ and $\sigma^{\pm 3}$ components \citep{chaisson75,hansen82} but
are not conclusive, since it is not clear that all spectral features
come from the same spatial region.  Nevertheless, the possibility that
$\sigma^{\pm 2}$ and $\sigma^{\pm 3}$ components may exist in OH
masers presents practical difficulties for observers of satellite-line
transitions, as noted by \citet{fish03} and \citet{hoffman05a}.
Conversion of the velocity difference of $\sigma$ components in a
Zeeman pair to a magnetic field strength is dependent upon the Zeeman
splitting coefficient, which is different depending on which $\sigma$
components are seen.  Traditionally it has been assumed that only the
$\sigma^{\pm 1}$ components are seen, for which a Zeeman splitting of
0.654 kHz~mG$^{-1}$ is appropriate at 1612 and 1720 MHz.  But it is
possible that several $\sigma$ components overlap for small Zeeman
splittings, in which case the Zeeman splitting coefficient appropriate
for conversion to a magnetic field strength may be a weighted average
of the splitting coefficients of the $\sigma$ components.  Indeed,
comparison of magnetic fields obtained from Zeeman splitting of the
1720 MHz transition are often a factor of 1.5 to 2 higher than those
obtained in the same spatial region at 1665 or 1667~MHz
\citep{fish03,caswell04}, although \citet{gray92} do note an instance
in W3(OH) in which the splitting between two 1720 MHz
features of opposite polarization appears consistent with their
interpretation as $\sigma^{\pm1}$ components with a splitting
coefficient (between $\sigma$ components) of 0.12~\kms~mG$^{-1}$.  It
has heretofore been unclear whether the \citeauthor{fish03}\ and
\citeauthor{caswell04} results indicate that 1720 MHz masers prefer
higher densities (which are correlated with magnetic field strength)
or that multiple $\sigma$ components from the same Zeeman group are
blended together.  It is interesting to note that the Zeeman splitting
coefficient between a blend of all $\sigma^+$ and $\sigma^-$
components in their LTE ratio of intensities is exactly twice the
splitting coefficient of the $\sigma^{+1}$ and $\sigma^{-1}$
components alone.

High spectral-resolution observations of OH masers are also important
in order to determine the maser lineshapes.  Theoretical models
suggest that maser lineshapes may be sensitive to the degree of
saturation and to the amount of velocity redistribution of molecules
along the maser amplification path
\citep{goldreich74,field94,elitzur98}.  While other masers have been
observed at high spectral resolution, such as 12 GHz CH$_3$OH masers
\citep{moscadelli03} and 22 GHz H$_2$O masers \citep{vlemmings05}, OH
masers have never been observed with both the high spectral resolution
required to determine the shape of the line wings and the high angular
resolution required to ensure that spatially-separated maser spots are
not blended together in the beam.  The lack of such observations may
be due to instrumental limitations.  Since the velocity extent of OH
maser emission in massive star-forming regions is typically several
tens of \kms, an appropriately wide bandwidth is usually selected in
order to observe all maser spots simultaneously.  Because the number
of spectral channels allowed by the correlator is generally limited
(e.g., 1024 channels per baseband channel for the Socorro correlator,
or only 128 channels in full-polarization mode), ground-state masers
are usually observed at 1~kHz (0.18~\kms) resolution to within a
factor of two.

It is in these interests that we have undertaken observations of OH
masers in two high-mass star-forming regions at very high spectral
resolution.  Orion KL was chosen in order to examine whether the
features observed by \citet{hansen82} consist of a single Zeeman group
or several spatially-unrelated maser features.  W3(OH) was chosen
because it is a frequently studied massive star-forming region with a
well understood magnetic field structure \citep{bloemhof92} and has
several bright maser features at 1612 and 1720 MHz
\citep{masheder94,argon00,wright04b}.

\section{Observations}
\label{observations}

The National Radio Astronomy Observatory\footnote{The National Radio
Astronomy Observatory is a facility of the National Science Foundation
operated under cooperative agreement by Associated Universities,
Inc.}'s Very Long Baseline Array (VLBA) was used to observe the
ground-state OH masers in two massive star-forming regions: W3(OH) and
Orion KL.  Data were collected starting at approximately 09 00 UT on
2005 September 20 using all 10 antennas.  Approximately 2.3 hours of
on-source observing time was devoted to W3(OH) and 1.0 hours to Orion
KL.  DA193 was also observed as a bandpass calibrator.

All four ground-state transitions (1612.23101, 1665.40184, 1667.35903,
and 1720.52998 MHz) were observed in dual circular polarization.  A
bandwidth of 62.5~kHz was divided into 512 spectral channels with
122~Hz channel spacing (0.02~\kms\ velocity spacing).  The usable
equivalent velocity bandwidth of about 10~\kms\ was centered at
$-44$~\kms\ LSR for W3(OH) and $+10$~\kms\ for Orion KL.  Many OH
maser spots fall outside this velocity range in Orion KL, but the
bandwidth was centered appropriately to include the 1612 MHz maser
feature at $+8$~\kms\ for which \citet{hansen82} claimed detection of
$\sigma^{\pm2}$ components.  The data were sampled in 4-level (2-bit)
mode.  A correlator averaging time of 4 seconds was used.  Due to the
large oversampling required to record 62.5~kHz of bandwidth, 122~Hz is
the highest spectral resolution available to the Socorro
correlator\footnote{The minimum sample rate is 2.0 Msamples s$^{-1}$.
With an oversampling factor of 16, the correlator playback interface
cannot accumulate a 2048-bit Nyquist-sampled FFT segment before
internal buffers are cleared automatically.}.  Four of the stations
(Brewster, North Liberty, Owens Valley, and St.~Croix) used the
original VLBA tape-based recording system, while the other six used
the newer Mark 5 disk-based recording system.

The data were reduced using the NRAO Astronomical Image Processing
System (AIPS).  Left circular polarization (LCP) data from the North
Liberty station were unusable due to anomalously low amplifier gain.
Data from the Hancock station were discarded due to strong radio
frequency interference (RFI).  Weaker RFI contaminated some data from
other stations as well.  An auto-correlation bandpass was applied to
the data (for further details see \S \ref{gradient}).  Each of the
four maser transition frequencies was self-calibrated and imaged
separately.  Additionally, the left and right circular polarizations
were self-calibrated and imaged separately from each other in the 1665
and 1667 MHz transitions, due to RFI that affected each polarization
differently.  No polarization calibration was applied, as the VLBA
polarization leakage is small enough for our scientific purposes.
Image cubes were created in both circular polarizations.  Each
velocity channel was searched for maser emission, and one or more
elliptical Gaussians were fitted to detected features using the
fitting routines of the AIPS task \texttt{JMFIT}.

\section{Results}
\label{results}

\begin{figure}[t]
\begin{center}
\resizebox{\hsize}{!}{\includegraphics{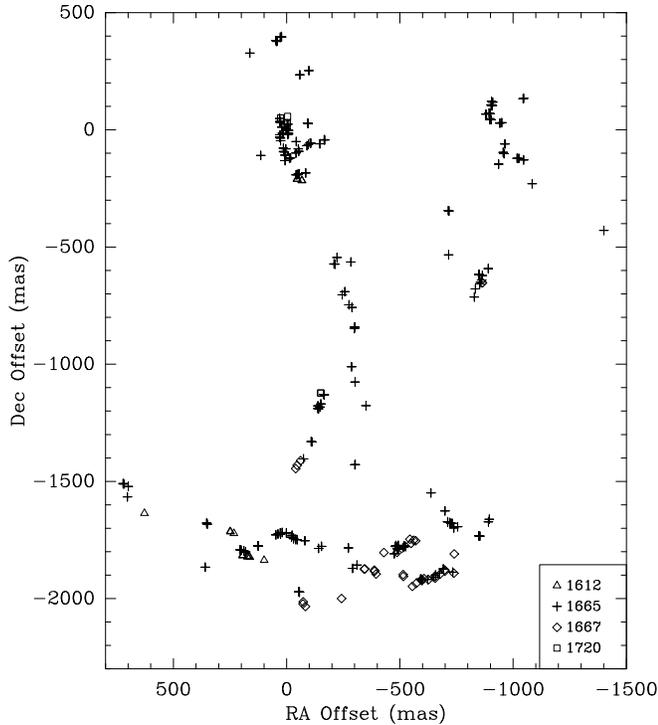}}
\end{center}
\caption{Ground-state OH masers in W3(OH).  Symbols indicate maser
  spots as follows: plus signs, 1665 MHz; diamonds, 1667 MHz;
  triangles, 1612 MHz; squares, 1720 MHz.
  \label{fig-map}}
\end{figure}

The detected spots are listed in Table \ref{tab-spots} and shown in
Figure \ref{fig-map}.  Symbols are plotted at the locations of peak
emission both in LCP and in RCP (right circular polarization).
Alignment of maser maps at different frequencies was accomplished by
comparison with the data of \citet{wright04b}.  We estimate that
resulting relative positional errors between frequencies is $\sim 10$
mas, due partly to errors in estimating spot positions in the two
epochs and partly to the inherent motion of the maser spots.  Maser
proper motions in W3(OH) are about 3 to 5~\kms\ \citep{bloemhof92},
which corresponds to 3 to 5 mas in the 9-year baseline between the
\citet{wright04a,wright04b} observations and the present data.  Zeeman
pairs are identified in Table \ref{tab-zeeman}.

Our map is qualitatively similar to the \citet{wright04b} map.  We
recover the vast majority of maser spots in their data.  Omissions may
be explained by the difference in sensitivity in the observations.
The \citet{wright04a,wright04b} observations spent nearly a factor of
5 more time on source with a factor of 4 coarser velocity resolution.
Additionally, data from the Hancock and (frequently) North Liberty
VLBA stations were not usable in the present observations, further
reducing our sensitivity.

We were unable to detect any satellite-line maser emission in Orion
KL.  Our pointing center was chosen to be 4\arcsec\ to the northeast
of the group marked ``Center'' in the map of \citet{johnston89}, which
is the probable location of the Stokes V spectrum interpreted as a
4-component Zeeman pattern centered at velocity 8.0~\kms\ by
\citet{hansen82}.  At this pointing center, peak amplitude loss to
time-average smearing is small (5\% at 7\arcsec).  Bandwidth smearing
is negligible with our extremely narrow spectral channels.  The
nondetection is therefore likely due to a decrease in the flux density
of the 8~\kms\ 1612 MHz features in the 20 years since the Johnston et
al.\ observations.  (Note that their measured flux densities are lower
than those of Hansen seven years prior.)  Because our primary goal in
these observations was to address the issue of satellite-line Zeeman
splitting and the Orion KL main-line data were of inferior quality to
the W3(OH) data, the Orion KL data were not further analyzed.

\subsection{Lineshapes and Gaussian Components}
\label{lineshapes}

\begin{figure}[t]
\begin{center}
\resizebox{\hsize}{!}{\includegraphics{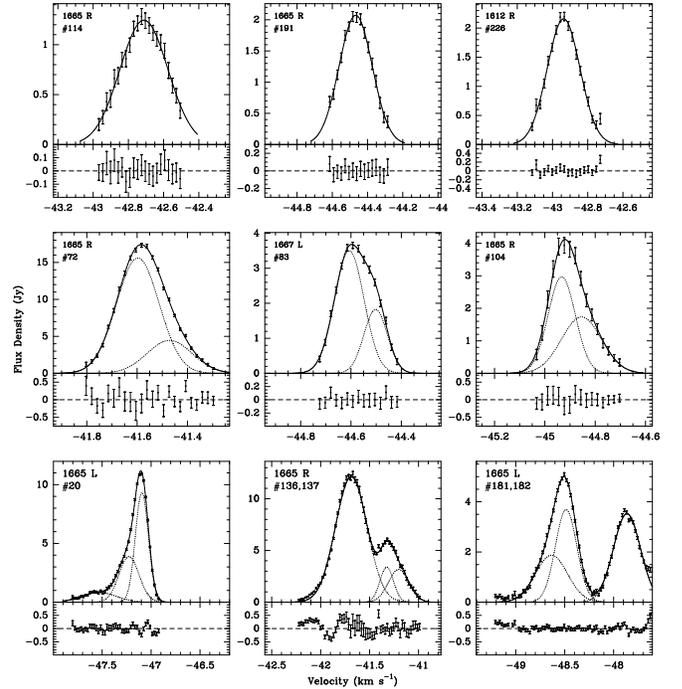}}
\end{center}
\caption{Spectra of selected features with a near Gaussian lineshape.
  The top portion of each panel shows the data and best Gaussian fit.
  The bottom portion shows the residuals.  Spot numbers as listed in
  Table \ref{tab-spots} are indicated for each panel.  The top,
  middle, and bottom rows show spectra fitted with one, two, and three
  Gaussian components, respectively.
  \label{fig-gaussians}}
\end{figure}

\begin{figure}[t]
\begin{center}
\resizebox{\hsize}{!}{\includegraphics{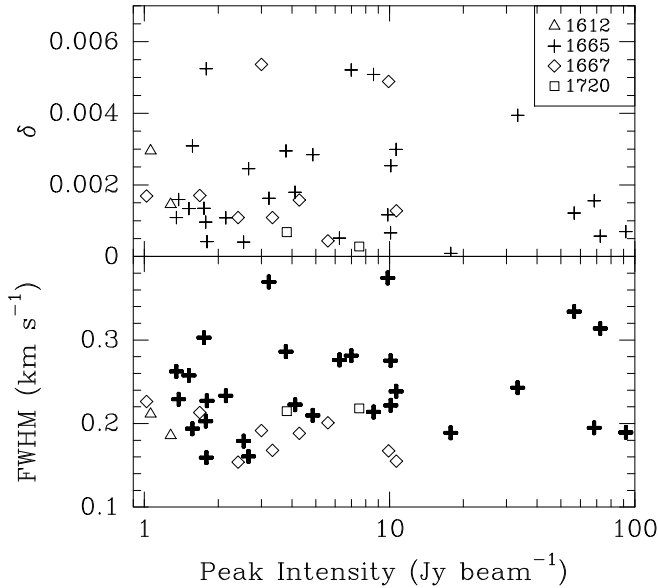}}
\caption{\emph{Top}: Normalized deviation from Gaussian shape for
  near-Gaussian maser features with peak intensity brighter than 1 Jy
  beam$^{-1}$.
  \emph{Bottom}: FWHM as a function of peak intensity.  Masers in the
  1665 MHz transition (bolded) appear to have a wider spectral profile
  than masers in the other ground state transitions.
  \label{fig-delta}}
\end{center}
\end{figure}

\begin{figure}[t]
\begin{center}
\resizebox{\hsize}{!}{\includegraphics{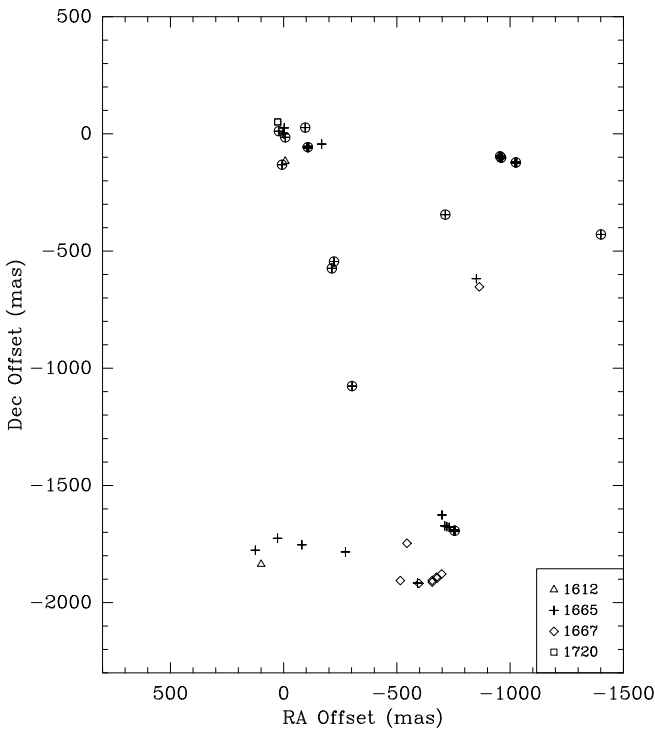}}
\caption{Map of spots included in analysis of lineshapes.
  Circled symbols indicate maser spots
  with fitted FWHM greater than 0.227~\kms; bolded symbols indicate maser
  spots with FWHM greater than 0.30~\kms.
  \label{fig-mapfwhm}}
\end{center}
\end{figure}

Even at this high spectral resolution, the OH maser spectral profiles
can usually be fitted well with one or a small number of Gaussian
components in the spectral domain.  The top panels in Figure
\ref{fig-gaussians} show selected maser spots with single-Gaussian
fits.  For spots weaker than $\sim 1$~Jy\,beam$^{-1}$ (about half the
spots), a single Gaussian component usually fits the spectral profile
due to signal-to-noise limitations.  For spots with a larger
signal-to-noise ratio, more complicated spectral profiles are seen as
well.  Some maser profiles appear skewed or have asymmetric tails.
For the most part, these can be fairly well fitted by two or three
Gaussian components, as shown in the middle and bottom panels of
Figure \ref{fig-gaussians}.  In some instances, two or more maser
lines at different velocities may appear at approximately the same
spatial location.  When multiple distinct peaks are present in the
spectral domain at the same spatial location, we identify each peak as
maser line for purposes of inclusion in Table \ref{tab-spots}.  It is
observationally cleanest to identify these as separate features,
although theoretical models indicate that under certain conditions the
spectrum of a single masing spot could be multiply peaked
\citep[e.g.,][]{nedoluha88,field94}.

The normalized deviation ($\delta$) of a lineshape from a Gaussian
shape can be defined by
\[
\delta = \frac{\int\left[I(v) - a_1 \exp(-v^2/a_2)\right]^2\,dv}
              {I_p^2 \, \Delta v},
\]
where $I(v)$ is the intensity distribution as a function of velocity,
$a_1$ and $a_2$ are parameters from the best Gaussian fit, $I_p$ is
the peak intensity, and $\Delta v$ is the FWHM of the distribution
\citep{watson02}.  Figure \ref{fig-delta} shows the distribution of
$\delta$ (calculated over the entire range of channels in which the
maser spot is detected) and the FWHM as a function of $I_p$ for masers
with a peak intensity greater than 1 Jy beam$^{-1}$.  Excluded from
consideration are masers with multiple peaks or extended asymmetric
tails (e.g., with profiles as in the bottom of Figure
\ref{fig-gaussians}) as well as maser features for which spatial
blending with a second maser feature within the beam prohibits
accurate determination of the spectral profiles of the two features
individually.

Derived values of $\delta$ range from 0 to $6 \times 10^{-3}$,
consistent with theoretical predictions by \citet{watson03}.  The
distribution is suggestive of a fall-off of $\delta$ for large values
of $I_p$, but with only five data points having $I_p > 30$ Jy
beam$^{-1}$, this result is not statistically significant.
\citeauthor{watson03} predict that $\delta$ attains a maximal value
when the stimulated emission rate, $R$, is a few times the pump loss
rate, $\Gamma$, and then decreases with increasing $R/\Gamma$ (i.e.,
as the maser becomes increasingly saturated)\footnote{In the saturated
regime, the maser intensity increases as a polynomial function of $R$
depending on the geometry \citep{goldreich72}, and thus $\log(I_p)
\propto \log(R/\Gamma)$.}.  Their model also predicts that when
$R/\Gamma$ is large enough for $\delta$ to decrease noticeably, the
line profile should rebroaden to nearly the thermal Doppler linewidth.
In our observations, the FWHM linewidth of maser spots has a range of
0.15 to 0.38 \kms\ and is independent of the peak intensity of the
maser spot.  This suggests that even the brightest masers in a typical
star-forming region (which are clearly saturated) may not be
sufficiently saturated to exhibit rebroadening.

The FWHM linewidth does appear to be a function of maser transition.
As shown in Figure \ref{fig-delta}, 1665 MHz masers are on average
broader than their other ground-state counterparts.  The mean and
sample standard deviation of the FWHM are $0.244 \pm 0.058$ \kms\ for
the 1665 MHz transition and $0.192 \pm 0.025$ \kms\ for the other
three transitions combined.  Sixteen of the 28 1665 MHz masers
brighter than 1 Jy beam$^{-1}$ are broader than 0.227 \kms, the
linewidth of the broadest maser in any of the other three transitions.
The spatial distribution of masers meeting the brightness and shape
criteria for the above analysis is shown in Figure \ref{fig-mapfwhm}.
As can be seen by comparison with Figure \ref{fig-map}, nearly all of
the broad 1665 MHz maser spots appear in regions where no ground-state
OH masers are found except in the 1665 MHz transition.  The only broad
1665 MHz masers found in proximity to masers of other ground-state
transitions are in the cluster of masers near the origin.  This region
is notable for the existence of highly-excited OH
\citep{baudry93,baudry98} and is presumed to mark the location of an
O-type star exciting the \hii\ region.  As such, it is likely that the
physical conditions change in this region over a shorter linear scale
than in other regions of W3(OH).

\subsection{Satellite-Line Splitting}
\label{satellite}

\begin{figure}[t]
\begin{center}
\resizebox{\hsize}{!}{\includegraphics{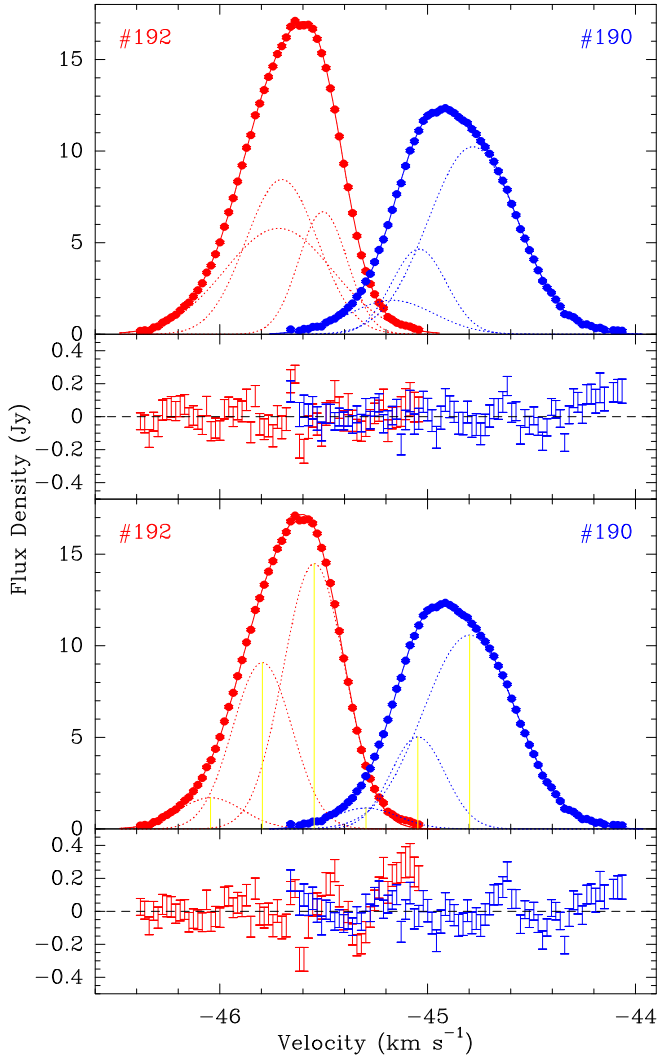}}
\caption{Fits of the brightest 1720 MHz Zeeman pair in W3(OH).
  \emph{Top}: Best three-component fit to each polarization.  LCP
  emission is shown in red and RCP in blue.  The individual fit
  components are shown as dotted lines, and the total fit spectra are
  shown as solid lines.  Residuals are shown below the fits.
  \emph{Bottom}: Best six-component fit constraining the center
  velocities of each component to be spaced as in Figure
  \ref{fig-intensities}.  Note that the pattern of intensities (in
  yellow) is not symmetric about a center velocity for the $\sigma^+$
  and $\sigma^-$ components.  The significantly poorer fit does not
  justify the reduction in the number of parameters.
  \label{fig-1720}}
\end{center}
\end{figure}

\begin{figure}[t]
\begin{center}
\resizebox{\hsize}{!}{\includegraphics{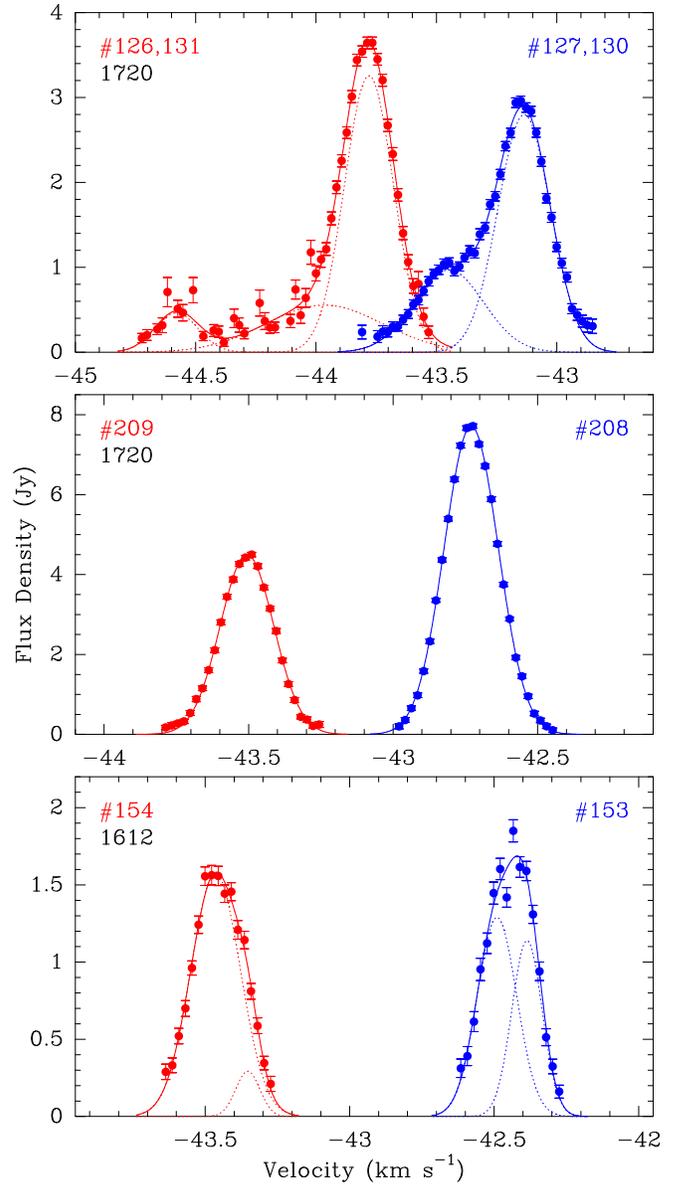}}
\caption{Spectra of satellite-line Zeeman pairs.  LCP emission is
  shown in red and RCP in blue.  Gaussian fits are superposed in solid
  lines; individual Gaussian components are shown in dotted lines.
  The top right and top left panels exhibit complicated lineshapes,
  the latter due to blending of several components within the beam.
  However, there is no evidence for $\sigma^{\pm2,3}$ components.
  \label{fig-satsplit}}
\end{center}
\end{figure}

There is no evidence of the presence of multiple $\sigma$ components
in the single-polarization spectra of the satellite-line (1612 and
1720 MHz) transitions.  We find four Zeeman pairs at 1720 MHz and
seven at 1612 MHz.  Figure \ref{fig-1720} shows the LCP and RCP
spectra of the brightest 1720 MHz Zeeman pair.  The top panels show
the best three-component fit to each polarization.  The residuals
suggest that a fourth, weak component may be required to fit the
high-velocity tail of RCP emission.  The magnetic field strength
derived from applying the splitting coefficient appropriate for
$\sigma^{\pm1}$ components to the velocities of the peak channels of
emission in spots 190 and 192 is $+6.6$~mG.

The velocities of the two strongest, narrow Gaussian components in
each circular polarization in the top panels of Figure
\ref{fig-1720} are consistent with a 2:1 splitting ratio centered
at approximately $-45.26$~\kms\ to within the errors in determining
the center velocities of the Gaussians.  The only lines in a
symmetric, incomplete Zeeman pattern with this ratio are the
$\sigma^{\pm1}$ and $\pi^\pm$ components.  Nevertheless, we reject the
possibility that the two brightest Gaussians correspond to $\pi$
components for several reasons.  First, they are only seen in one
circular polarization, while $\pi$ components should be 100\% linearly
polarized (although see \citealt{fish06} for a discussion of the
possibility of $\pi$ components with nonzero circular polarization
fractions).  Second, the $\pi^0$ component is missing from this
pattern, although theory predicts that it should be stronger than
$\pi^\pm$ components.  Third, no other $\pi$ components are seen in
the ground-state masers in W3(OH) \citep{garciabarreto88}.  Linear
polarization is rare in W3(OH); all masers are more circularly
polarized than linearly polarized.

The bottom panels show the best fit constraining the center velocities
of each component to be in the ratio expected from the Zeeman pattern
of multiple $\sigma$ components in each polarization, as shown in
Figure \ref{fig-intensities}.  It is not the case that the two
Gaussian components closest to the systemic velocity are brightest.
This argues against interpretation of the spectra as $\sigma^{\pm
1,2,3}$ components in their LTE ratios.  It is more probable that the
same factors that produce asymmetric, non-Gaussian lineshapes in the
main-line transitions (as in the middle and bottom panels of Figures
\ref{fig-gaussians}) also produce non-Gaussian asymmetries in the
$\sigma^{\pm 1}$ components at 1720 MHz.  Indeed, at higher angular
resolution, \citet{masheder94} note that features 190 and 192 are each
actually a cluster of several maser spots.  This is consistent with
the increasing intensities in each polarization toward higher
velocity, suggesting that our observed features may be the result of
blending of (at least) three nearby Zeeman pairs with a regular shift
in velocity but approximately the same magnetic field (coincidentally
also $+6.6$~mG if interpreted as $\sigma^{\pm 1}$ components).  This
magnetic field value is consistent with the two nearby Zeeman pairs to
the northeast: $+6.4$~mG from spots 212 and 218 at 1665 MHz and
$+6.8$~mG from spots 208 and 209 at 1720 MHz.  (Note that this latter
Zeeman pair, shown in the middle panel of Figure \ref{fig-satsplit},
is unquestionably comprised solely of $\sigma^{\pm1}$ components,
since there is only one feature in each circular polarization and
interpretation of these features as $\sigma^{\pm2}$ components would
imply a magnetic field strength of $+2.2$~mG, a value too small to
be consistent with the 1665 MHz magnetic field or any other magnetic
field strength in the cluster of maser spots near the origin.)

Figure \ref{fig-satsplit} shows the LCP and RCP spectra of the other
two 1720 MHz Zeeman pairs and one 1612 MHz Zeeman pair in W3(OH).  The
multiple peaks in the single-polarization spectra of the top panel are
again due to blending of two adjacent maser spots.  It is clear that
these are not due to spatially-shifted $\sigma^{\pm2,3}$ components
from a single Zeeman pattern, since the spectra are not symmetric by
reflection across a single, systemic velocity.  We interpret the
spectra as two Zeeman pairs, each with a different magnetic field
strength.  Asymmetric amplification of the various peaks in LCP and
RCP may also be partly due to the large velocity range spanned ---
greater than 1.4~\kms\ from the low-velocity peak in LCP to the
high-velocity peak in RCP.  Since this is more than twice the
turbulent velocity dispersion of a maser cluster in W3(OH)
\citep{reid80}, it would be expected that the emission from multiple
maser spots in this velocity range might be amplified by different
amounts.

The middle panel in Figure \ref{fig-satsplit} shows a 1720 MHz Zeeman
pair that is well fitted by a single Gaussian component in each
polarization.  The bottom shows a 1612 MHz Zeeman pair.  It is clear
from the velocities of the fit components that the lines are not
produced from multiple $\sigma$ components of a single Zeeman pattern.
Emission from other masers in the 1612 MHz transition is qualitatively
similar to these Zeeman pairs.  The image cubes of 1612 and 1720 MHz
emission were searched thoroughly at the locations of the detected
masers for indications of weak emission at other velocities.  No
emission was detected to within the limits of our noise except as
listed in Table \ref{tab-spots}.

\subsection{Positional Gradients}
\label{gradient}

\begin{figure*}[t]
\begin{center}
\resizebox{\hsize}{!}{\includegraphics{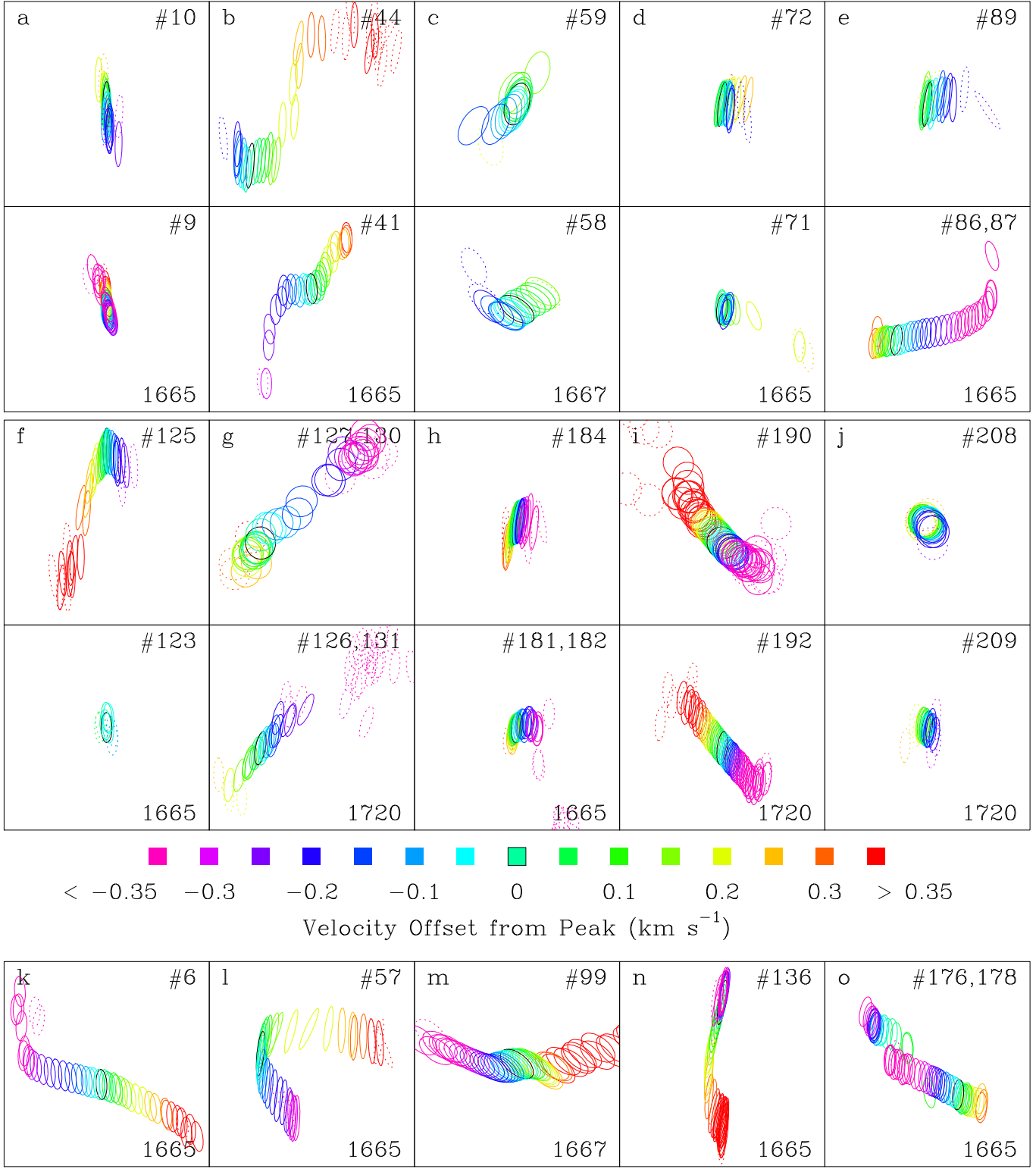}}
\caption{Positional gradients as a function of LSR velocity for a
  selection of maser spots.  Boxes are 20 mas on a side.  The position
  in each velocity channel is represented as an ellipse one-tenth the
  size of the undeconvolved spot size, fit as an elliptical Gaussian.
  The black contour denotes the channel of peak emission; colors
  indicate velocity-shifted channels.  Dotted contours indicate
  channels in which the SNR is less than 10.  Panels a through j show
  the gradients in ten Zeeman pairs.  Boxes are aligned separately for
  the RCP (always on top) and LCP (bottom) data.  The gradients of the
  RCP and LCP components of the three Zeeman pairs at 1720 MHz are
  similar in amplitude and well aligned in position angle.  Panels k
  through o show additional interesting spots.  The velocity gradient
  measured in \kms~mas$^{-1}$ is inversely proportional to the
  positional gradient (presented here) measured in mas~(\kms)$^{-1}$.
  \label{fig-gradients}}
\end{center}
\end{figure*}

\begin{figure}[t]
\begin{center}
\resizebox{\hsize}{!}{\includegraphics{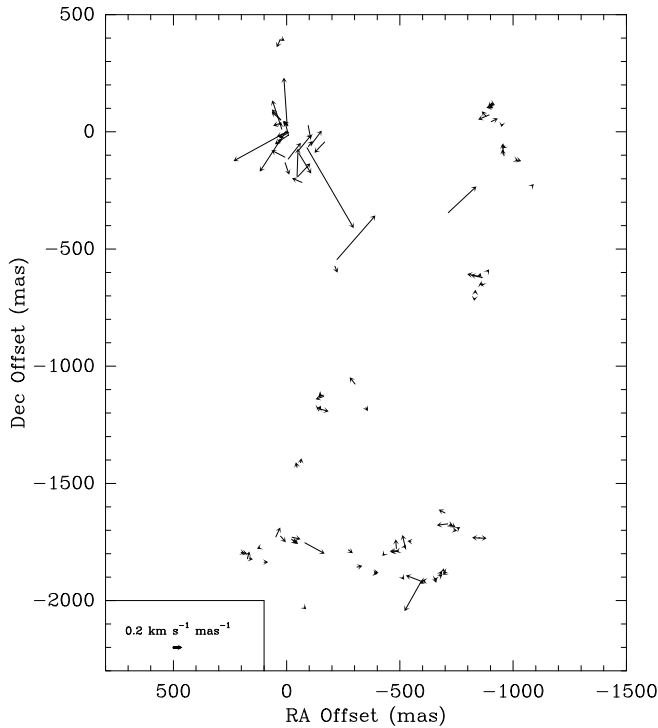}}
\caption{Map of velocity gradients of spots with peak brightness
  greater than 1 Jy beam$^{-1}$.  Arrows point in the direction of
  change with increasing line-of-sight velocity.  Velocity gradients
  are generally large in the cluster near the origin and smaller
  elsewhere.
  \label{fig-gradientmap}}
\end{center}
\end{figure}

\begin{figure}[t]
\begin{center}
\resizebox{\hsize}{!}{\includegraphics{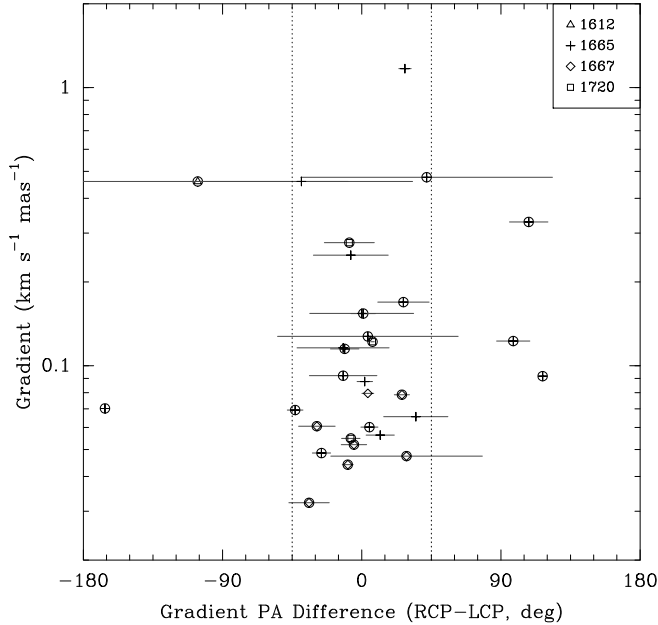}}
\caption{Velocity gradients and position angle differences of those
  velocity gradients for components of Zeeman pairs (circled) and
  echoes (not circled; defined in \S \ref{gradient}).  Error bars are
  shown along the abscissa only.  The magnitude of the velocity
  gradient is determined by taking the inverse of the average of the
  positional gradients as measured in mas~(\kms)$^{-1}$ of the RCP and
  LCP components.  Dotted lines indicate RCP$-$LCP position angle
  differences of $\pm$45\degr.  In most Zeeman pairs, LCP and RCP
  gradients are roughly aligned.  This is especially true for the
  three 1720 MHz Zeeman pairs.
  \label{fig-zeemangrads}}
\end{center}
\end{figure}

\begin{figure}[t]
\begin{center}
\resizebox{\hsize}{!}{\includegraphics{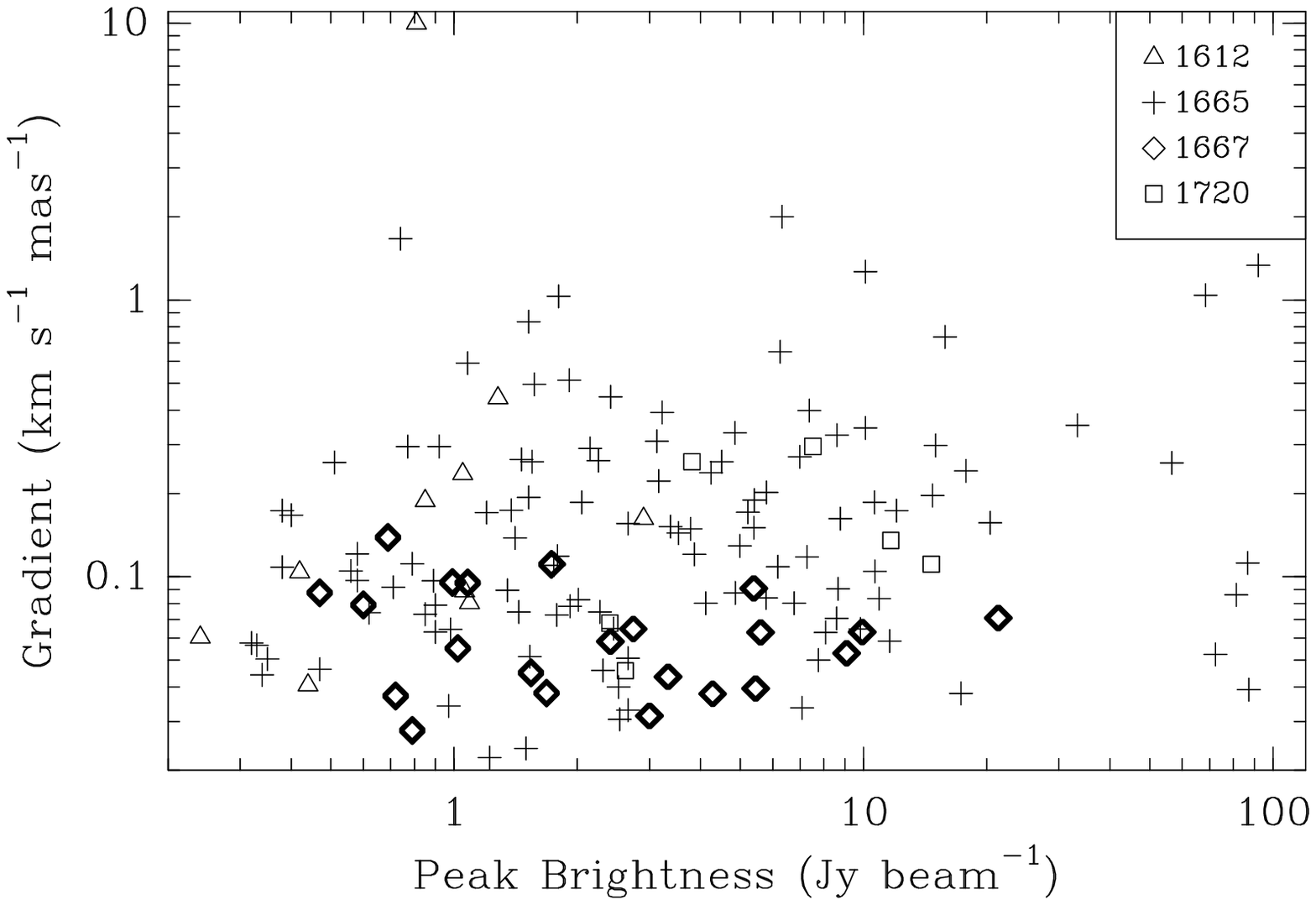}}
\caption{Magnitude of velocity gradient as a function of peak
  brightness.
  There does not appear to be a correlation between the velocity
  gradient and maser brightness.  However, of the 22 1667 MHz maser
  spots for which a gradient is determined (displayed in bold for
  contrast), the largest gradient is 0.14~\kms~mas$^{-1}$.
  \label{fig-gradflux}}
\end{center}
\end{figure}

In general, the position of a maser spot is seen to vary across the
linewidth \citep[e.g.,][]{moscadelli03,hoffman03}.  Figure
\ref{fig-gradients} shows the maser position as a function of LSR
velocity for a sample of maser spots.  The position of the center of
the best-fitting elliptical Gaussian usually varies linearly as a
function of velocity.  In some instances the position may trace out a
curving structure rather than a straight line, but all maser spots
display organization in their position as a function of frequency.

Table \ref{tab-spots} includes the velocity gradient and position
angle (degrees east of north) in the direction of increasing velocity
for each maser spot.  The velocity gradients were determined
algorithmically.  On both sides of the peak, the nearest channel with
emission below half of the peak brightness was identified.  The
velocity difference between these two channels was divided by the
difference in positions.  For a Gaussian spectral profile this
corresponds to dividing the FWHM by the difference of the positions
across the FWHM, but it is algorithmically implementable for any
emission spectrum, including spectra with multiple peaks, as in the
middle and bottom panels of Figure \ref{fig-gaussians}.  For
consistency with \citet{moscadelli03}, we report the velocity
gradients in units of \kms~mas$^{-1}$ rather than the positional
gradient in mas~(\kms)$^{-1}$.  A large positional gradient
corresponds to a small velocity gradient, and vice versa.

These gradients appear to be real, not an artifact due to residual
calibration or bandpass phase errors.  Comparison of selected bright
maser spots in different regions of W3(OH) indicate that positional
gradients determined from applying the auto-correlation (real)
bandpass are consistent with those determined from applying the
cross-correlations (complex) bandpass to within measurement errors.
Since the auto-correlation bandpass has a higher signal-to-noise ratio
and the phases of the cross-correlation bandpass are constant with
frequency over the region of interest, the auto-correlation bandpass
was applied.  In addition, combinations of plots of the Right
Ascension or Declination positional gradients versus Right Ascension
offset, Declination offset, or LSR velocity are all consistent with a
random scatter about zero (as with Figure 6 in
\citeauthor{moscadelli03}), both for individual transitions and
polarizations as well as for the ensemble of all maser spots with
detected positional gradients as listed in Table \ref{tab-spots}.
However, the two-dimensional distribution does show larger velocity
gradients (smaller positional gradients) near the origin (Figure
\ref{fig-gradientmap}).  Note that the origin is \emph{not} near the
location of the reference spots for self-calibration except at 1720
MHz and the LCP polarization at 1665 MHz, nor is it near the pointing
and correlation center \citep[taken from][]{argon00}, which is at
$(\Delta\alpha,\Delta\delta) \approx (-884,+311)$ mas.

It is probable that some gradients are the result of two maser spots
within a beamwidth that blend together spectrally.  One clear instance
of this is shown in the top panel of Figure \ref{fig-satsplit}.  Only
one feature is detected in each circular polarization in each spectral
channel.  Yet it is clear from the spectra that there are at least two
distinct maser spots in each polarization.  The weaker peak is to the
northwest of the strong peak (panel g of Figure \ref{fig-gradients}).
The centroid of the fitted Gaussian is effectively a weighted average
of the two positions at velocities intermediate to the two peak
velocities.  This effect is more prominent in RCP due to the smaller
velocity offset between the two peaks.  Nevertheless, there is a real
positional gradient associated with each of the maser spots as well,
as is clearest in the uncontaminated blue wing of the bright features.

Velocity gradients of the RCP and LCP components of a Zeeman pair are
generally aligned.  Figure \ref{fig-zeemangrads} shows the
distribution of position angle differences between the velocity
gradients of the RCP and LCP components of Zeeman pairs.  These
position angle differences are also shown for ``echoes,'' i.e.,
spectral features detected in the opposite circular polarization and
same location and line-of-sight velocity as another strong, partially
linearly polarized spectral feature due to the fact that both circular
feeds of a telescope are sensitive to linear polarization.  These
detections are not a result of telescope polarization leakage; in most
cases, the brightness of the weaker polarization feature is more than
25\% of that of the stronger polarization feature, while polarization
leakage of the VLBA feeds is only 2 to 3\% \citep{wrobel05}.  Since an
echo is a second, weaker detection of a single maser spot, both a
maser spot and its echo would be expected to have essentially the same
gradient.  We find this to be the case; for the 9 maser spots for
which a gradient can be determined algorithmically both for itself and
its echo, all have gradient polarization angle differences less than
40\degr.  Of the 23 Zeeman pairs for which gradients can be obtained
for both components, the RCP and LCP components are aligned to within
better than 45\degr\ for 18 of them.  Larger deviations for the other
pairs can usually be attributed to spatial blending with nearby maser
spots.  The alignment of RCP and LCP gradients is especially
pronounced in the bright Zeeman pairs at 1720 MHz.  Their spectra and
positions are shown in Figures \ref{fig-1720}, \ref{fig-satsplit}, and
\ref{fig-gradients}.  In each case, there is a clear, linear
positional gradient that is similar for both components of the Zeeman
pair.

The magnitude of the velocity gradient of a maser spot does not
display a clear correlation with its peak brightness, as shown in
Figure \ref{fig-gradflux}.  However, there does appear to be an
absence of 1667 MHz maser spots with large velocity gradients.  (That
is, 1667 MHz masers appear to have large \emph{positional} gradients
as a function of line-of-sight velocity.)  This is consistent with
observations by \citet{ramachandran06}.  It is unclear whether the
line-of-sight velocity gradient projected onto the plane of the sky
necessarily allows inference of the line-of-sight velocity gradient
along the amplification path.  Large velocity gradients along the
amplification length may destroy the velocity coherence required for
significant amplification, so the population of detectable maser spots
may have an inherent bias in favor of areas where the projection of
the velocity gradient along the line of sight is small.  But in \S
\ref{transitions} we present further evidence that the velocity
gradient along the amplification path is indeed small in 1667 MHz
masers.

There does not appear to exist a correlation between the orientation
of the gradient of a maser spot and its proper motion vector.  From
the list of 1665 MHz maser spots for which \citet{bloemhof92} were
able to measure a proper motion, approximately three dozen spots with
measurable positional gradients were recovered in our observations.
Since the Bloemhof et al.\ data were not phase referenced, multiple
reference frames consisting of their proper motions with an added
constant vector were compared against our positional gradient vectors.
No clear correlations were found.  Proper motion maps of the OH masers
in W3(OH) display a clear large-scale pattern of motions
\citep{bloemhof92,wright04a}, while the map of gradients shows no such
large-scale organization, with the possible exception of the cluster
near the origin (Figure \ref{fig-gradientmap}), where velocity
gradients are large (i.e., positional gradients are small).  If there
is a connection between observed maser velocity gradients and material
motions, it is probable that it is the turbulent motions that
dominate, not the large-scale organized motions.

Likewise, the gradients do not correlate with linear polarization
fraction (which is zero for most maser spots) or polarization position
angle, as determined from \citet{garciabarreto88}.  The magnetic field
direction can theoretically be derived from the linear polarization
fraction and position angle \citep[e.g.,][]{goldreich73a}, although
empirical data suggest that recovery of the full, three-dimensional
orientation of the magnetic field may not actually be possible at OH
maser sites \citep{fish06}.

\subsection{Deconvolved Sizes and Maser Geometry}
\label{deconv}

\begin{figure}[t]
\begin{center}
\resizebox{\hsize}{!}{\includegraphics{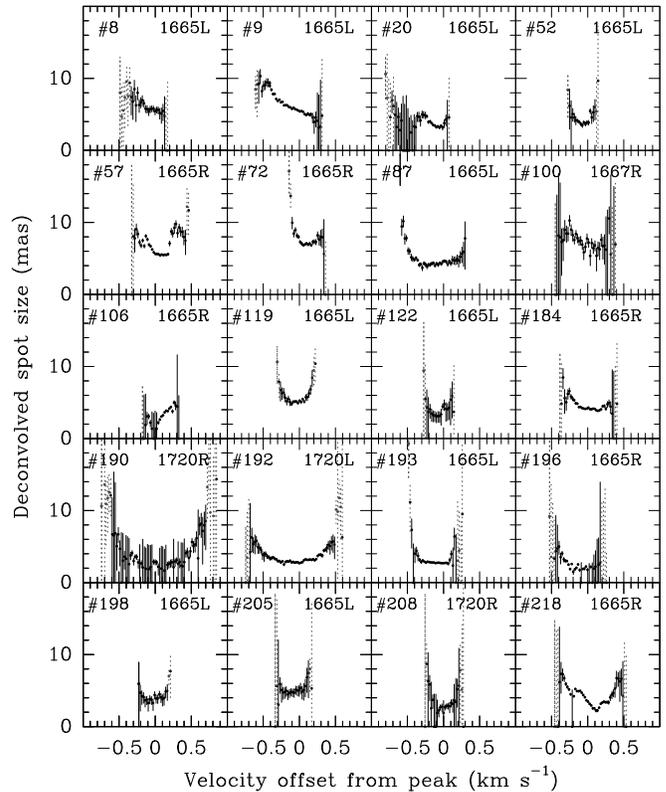}}
\end{center}
\caption{Deconvolved spot sizes as a function of velocity for selected
  maser spots, identified by spot number and transition.  The
  geometric mean of the deconvolved major and minor axes is plotted as
  a function of velocity offset from the channel of peak flux density.
  Dots represent the nominal deconvolved spot size, while the lines
  represent the allowed range of values within the noise.  Grey,
  dotted lines represent measurements with an SNR of less than 10.
\label{fig-deconv}}
\end{figure}

The apparent size of a maser may be a function of frequency offset
from line center, due to saturation effects dependent on the maser
geometry.  For example, \citet{elitzur90} calculates that the size of
a spherical maser should increase exponentially with $|\nu -
\nu_0|/\Delta\nu_\mathrm{D}$, where $\nu_0$ is the line center
frequency and $\Delta\nu_\mathrm{D}$ is the Doppler linewidth.  This
effect can be large; Elitzur calculates that the apparent spot size at
half the Doppler width may be twice that at line center (peak flux).

Figure \ref{fig-deconv} shows deconvolved spot sizes as a function of
velocity offset from the channel of peak emission for 20 selected
maser spots.  Displayed masers were selected under the criteria that
they have a peak flux density of at least 7 Jy and not have obvious
spatial blending with other maser emission.  Minimum nominal
deconvolved spot sizes typically range from 3 to 6 mas, consistent
with results obtained for 1665 MHz masers by \citet{garciabarreto88},
although the apparent sizes of 1720 MHz masers are much bigger than
the $\leq 1.2$~mas upper limit obtained by \citet{masheder94} (see \S
\ref{satellite} for discussion of probable spatial blending in spot
numbers 190 and 192).  In general, maser spot sizes appear to increase
toward the line wings, although the degree to which the spot size
increases with frequency offset from center (or indeed whether it does
at all) is different with each maser spot.  In some spots, the spot
size is a complicated function of frequency.  It is possible that some
maser spots display additional structure on scales smaller than the
beam size, which could cause the spot size to be overestimated over
part or all of the line profile.  In any case, the variation of spot
size over the observable line profile is sufficiently small and
variable to preclude accurate determination of the functional form of
the apparent spot size as a function of frequency (and therefore
geometry).

The velocity offset at which the maser spot size doubles is generally
greater than 0.5~\kms.  For a spherical maser, this implies a Doppler
width greater than 1.0~\kms, based on the \citet{elitzur90} model,
which would require a kinetic temperature in excess of 400~K.  This
value is more than a factor of two higher than the inferred effective
temperature of the ambient radiation field \citep{walmsley86}.  It is
probable that the geometry of the OH masers in W3(OH) is not
spherical.  Other theoretical considerations lead \citet{goldreich72}
to conclude that a filamentary geometry is more typical of
astrophysical masers.  Alternatively, several different (spherical)
clumps may overlap along the line of sight to produce a detectable
maser.

\section{Discussion}
\label{discussion}

\subsection{$\sigma$ Components in Satellite-Line Transitions}
\label{sigmas}

We find no evidence of the presence of $\sigma^{\pm2,3}$ components in
the 1612 and 1720 MHz satellite lines of OH.  Some of the spectral
profiles in the 1720 MHz transition appear to consist of several
Gaussian components (Figures \ref{fig-1720} and \ref{fig-satsplit}),
but the velocities and intensities of these components are not
consistent with what is expected by theory (Figure
\ref{fig-intensities}).  The nondetection of $\sigma^{\pm2,3}$
components lends support to the prediction that cross-relaxation
across magnetic sublevels will favor amplification of the
$\sigma^{\pm1}$ components over the other $\sigma$ components
\citep{goldreich73b}.  For the brightest 1720 MHz maser, our
nondetection of accompanying maser components requires that
$S_{\sigma^{\pm1}}/S_{\sigma^{\pm2}} >$ a few hundred.  Future
observations of sources with stronger satellite-line OH maser
emission, such as G43.165$-$0.028 \citep{argon00} and G331.512$-$0.103
\citep{caswell99}, could improve on this by more than a factor of 10.

Could $\sigma^{\pm2}$ components ever be observed in a maser source?
The number of gain lengths for a $\sigma^{\pm1}$ component will be
twice that of a $\sigma^{\pm2}$ component over the same physical
region of space.  For unsaturated amplification, the intensity depends
exponentially on the number of gain lengths, effectively prohibiting
detection of the $\sigma^{\pm2}$ components.  (Since the number of
gain lengths for the $\sigma^{\pm1}$ components is $\gtrsim 20$
\citep{goldreich73b}, the $\sigma^{\pm2}$ components would be weaker
by a factor of $\gtrsim e^{10}$).  If the $\sigma^{\pm1}$ components
are highly saturated, it is possible that the $\sigma^{\pm2}$
components would be detectable, provided that the populations of the
magnetic sublevels are not redistributed by radiative transitions
connecting these levels with the far infrared.  It should be noted
that two different radiative effects are likely operating in
satellite-line masers.  First, cross-relaxation of the magnetic
sublevel populations due to radiative transitions connecting these
levels with the far infrared will favor the $\sigma^{\pm1}$ components
\citep{goldreich73b}.  This effect can operate over frequency
differences much greater than the linewidth of a single maser
component.  Second, velocity redistribution inherent in a
three-dimensional geometry causes maser lines to remain narrow even
during saturation, preventing rebroadening to the Doppler width
\citep{field94}.  Velocity redistribution causes flux from the
linewings to move toward the line center, effectively changing the
frequency on the order of a maser linewidth.  Cross-relaxation of
populations of the magnetic sublevels will be unavoidable in regions
of strong infrared radiation.  Thus, it is probable that
$\sigma^{\pm2}$ components will not be detectable in massive
star-forming regions.

It is also likely that 1720 MHz supernova remnant masers will not
display evidence of $\sigma^{\pm2}$ components.  Supernova remnant
OH masers are collisionally pumped \citep{elitzur76,frail94}, so it
may be possible to avoid far infrared cross-relaxation among magnetic
sublevels.  However, the Zeeman splitting is usually less than the
maser linewidth \citep[e.g.,][]{hoffman05a,hoffman05b}, resulting in
blending of multiple maser components into a single maser line.
Velocity redistribution would likely destroy the signature of
$\sigma^{\pm2,3}$ components, if emission in these modes is produced.

As mentioned in \S \ref{introduction}, if $\sigma^{\pm2,3}$ components
are not blended with $\sigma^{\pm1}$ components at 1720 MHz, there is
observational evidence that the magnetic field, and hence the density,
at sites of 1720 MHz maser emission in massive star-forming regions
may be higher than at sites of 1665 and 1667 MHz maser emission
\citep{fish03,caswell04}.  Our data indicate that $\sigma^{\pm2,3}$
components, if they exist, are so weak as to have no effect on the
observed emission.  Thus, the value of the magnetic field obtained
from assuming a Zeeman splitting coefficient appropriate for pure
$\sigma^{\pm 1}$ components is reliable.  Using this coefficient, the
three brightest 1720 MHz Zeeman pairs in W3(OH) are consistent with
the magnetic field strengths derived from nearby main-line Zeeman
pairs.  Models of \citet{pavlakis96} suggest that 1720 MHz maser
activity may be favored at densities near or just above those for
which 1665 MHz maser activity occurs.  Thus, in an ensemble of OH
maser sources, it would be expected that the magnetic fields derived
from 1720 MHz Zeeman splitting would be skewed higher than those
obtained at 1665 MHz (consistent with the findings of \citealt{fish03}
and \citealt{caswell04}) , although the magnetic field strengths
derived at 1665 MHz and 1720 MHz would be similar in some of those
sources (consistent with this work).

\subsection{Comparison of Maser Transitions}
\label{transitions}

Our results for the properties of hydroxyl masers at high spectral
resolution are remarkably similar to those found in a similar study of
12.2 GHz methanol masers in W3(OH) \citep{moscadelli03}.  FWHM line
widths of single-Gaussian fits range from 0.15 to 0.38~\kms\ in OH, as
compared with the range 0.14 to 0.32~\kms\ in CH$_3$OH.  Normalized
deviations from a Gaussian shape are several $\times 10^{-3}$ for both
the OH and CH$_3$OH masers.  Gradients in the spot position as a
function of velocity are observed in both species, with similar
amplitudes.  The OH masers in our sample have velocity gradients as a
function of position (i.e., the inverse of a positional gradient as a
function of line-of-sight velocity) of 0.01 to 1~\kms~AU$^{-1}$ (with
one outlier at 5~\kms~AU$^{-1}$), as compared to 0.02 to
0.30~\kms~AU$^{-1}$ in a smaller sample of CH$_3$OH masers
\citep{moscadelli03}.  The similar observational characteristics of OH
and CH$_3$OH masers are not surprising given that these molecules form
in the same environment \citep{hartquist95}, are both excited under
similar conditions \citep{cragg02}, and appear in close proximity
\citep{etoka05}.

Since the ground-state transitions of OH have large Zeeman splitting
coefficients, an apparent velocity gradient could be the result of a
magnetic field gradient.  Indeed, in the cluster of maser spots
located near the origin in Figure \ref{fig-map}, the line-of-sight
velocity gradients as a function of position are large, and the
magnetic field strengths are large and change significantly on a small
spatial scale \citep[see Table \ref{tab-zeeman} of the present work as
well as Figure 13 of][]{wright04b}.  Likewise, the velocity gradients
are small in the cluster of masers near $(\Delta\alpha,\Delta\delta) =
(-800,-700)$~mas, where the magnetic field strengths are small and the
gradient of the magnetic field as a function of position is small.
But the observed velocity gradients cannot be entirely due to magnetic
field gradients, since they are also observed in methanol masers
\citep{moscadelli03}, in which Zeeman splitting is negligible.

In the absence of velocity redistribution between velocity subgroups
in the masing region, the linewidth of a saturated maser will in
general increase as the amplification (and hence, intensity) of the
maser increases \citep{goldreich74}, although maser linewidths remain
narrow even during saturated amplification when trapped infrared
radiation is included in the theory.  The lack of single-Gaussian
lineshapes with FWHM greater than 0.4~\kms\ combined with the absence
of a correlation between FWHM linewidth and maser flux density
suggests that line rebroadening does not occur, even for the brightest
OH masers.  \citet{field94} suggest that velocity redistribution is
important at 1665 MHz, which would produce narrow, single-peaked maser
lines, as observed.  In extreme cases, increasing amplification may
cause the line center to go into absorption, resulting in two very
narrow maser lines at different velocities \citep{gray91,field94}.
The addition of a velocity gradient over the amplification length can
also produce very strong, leptokurtic intensity profiles.  However,
large velocity gradients in the presence of complete velocity
redistribution can also produce multiply-peaked spectral profiles, which
we do not observe.  While spectral profiles do sometimes exhibit more
than one peak (as in Figure \ref{fig-gaussians}), it is neither the
case that the individual peaks in the spectrum are abnormally narrow
nor that the overall spectrum resembles a single broad Gaussian whose
center is strongly absorbed.  It is more probable that these spectra
are indicative of two or more spatially distinct maser spots
blended within a beamwidth.  VLBI studies of other sources find that
it is common for several distinct maser spots to be found within
several milliarcseconds of each other \citep[e.g.,][]{slysh01,fish05}.

We find that the linewidths of 1665 MHz masers are greater than the
other ground-state OH masers.  While 1665 MHz masers are usually the
brightest OH masers in a source, the lack of a correlation between the
linewidth and maser intensity indicates that saturated rebroadening is
not the cause of the larger linewidths at 1665 MHz.  One possible
explanation may involve the large Zeeman splitting coefficient at 1665
MHz.  A magnetic field gradient of 0.34 mG is sufficient to shift the
center velocity of a 1665 MHz maser by the 0.2~\kms\ FWHM typical in
other transitions; necessary magnetic fields for similar shifts are
0.56 mG at 1667 MHz and over 1.6 mG in the satellite-line transitions.
Observations of a larger sample of interstellar maser sources suggest
that the magnetic field strength typically varies by a few tenths to a
full milligauss in a typical cluster (projected dimension of several
$\times 10^{15}$~cm) of maser spots \citep{fish06}.  Since the
amplification length (along the line of sight) is likely a factor of a
few smaller than the clustering scale, it is reasonable that the
magnetic field strength may change by a few tenths of a milligauss
over the amplification length.  If so, and if velocity redistribution
is not total, it is possible that the resulting spectral profile would
be broader.  Under these assumptions, it would be expected that
broader 1665 MHz masers would appear in regions where the gradient of
the magnetic field is large.  The central cluster does contain several
broad 1665 MHz maser spots, and it is clear that the magnetic field
strength varies significantly over a small spatial scale in this
region.  However, the other broad OH masers in W3(OH) appear in
regions where the magnetic field strength is sampled (in this study
and in \citealt{wright04a,wright04b}) by only one or a few Zeeman
pairs, so it is difficult to obtain an estimate of the gradient of the
magnetic field in these locations.  Indeed, since regions of large
magnetic field gradients may not favor amplification of both
$\sigma$-components in a Zeeman pair \citep{cook66}, it is possible
that the magnetic field gradients in these regions are large.

A related possiblity is that the broad 1665 MHz masers sample a region
of parameter space in which amplification of only the 1665 MHz masers
is favored.  Due to Zeeman splitting, a large magnetic field gradient
might be expected to act akin to a large velocity gradient, although
rigorous theoretical examination of the effect of magnetic field
gradients in maser sites is lacking.  With the exception of the
central cluster of maser spots, in which physical conditions likely
change substantially over a small spatial scale, all other broad 1665
MHz masers appear in regions where the only ground-state OH masers
found are 1665 MHz masers.  Models by \citet{pavlakis96} indicate that
for radiatively-pumped OH masers, amplification of 1667 MHz decreases
significantly as the velocity gradient over the amplification path
increases from 1~\kms\ to 2~\kms, while 1665 MHz maser amplification
remains relatively unaffected.  This is in excellent qualitative
agreement with \citet{gray92}, who find that amplification of the 1667
MHz transition falls off with increasing velocity shift.  In our
observations, no broad 1665 MHz maser is found in the vicinity of 1667
MHz masers; in fact, 1667 MHz masers are the only ground-state
transition absent from the highly active central cluster of masers.
These facts fit well with the observation that 1667 MHz masers have
small line-of-sight velocity gradients in the plane of the sky (see
Figure \ref{fig-gradflux}), suggesting that the gradient of the
line-of-sight velocity along the amplification path may be small as
well.

It should be noted that the central cluster of masers also includes a
4765 MHz maser \citep{gray01}, for which inversion requires a small
velocity gradient \citep{pavlakis96}.  However, this maser is near the
southern edge of the cluster \citep{etoka05}, where the magnetic field
gradient is small.  It may be the case that even in clusters with
large velocity gradients, subregions exist in which the velocity
gradient along the line of sight is small.  In any case, it is not yet
established whether gradients in the centroid of a maser spot as a
function of line-of-sight velocity also provide information as to the
line-of-sight velocity distribution along the amplification path of a
maser spot.  If maser amplification is only favored for a narrow range
of velocity gradients along the amplification path, an unavoidable
observational bias will exist.  But velocity redistribution may weaken
the correlation between velocity gradients and maser gain.  Further
theoretical and observational work may be required to resolve these
issues.

\section{Conclusions}

We have observed over 250 ground-state OH maser spots at very high
spectral resolution.  Spectral profiles are generally well fit by one
or a small number of Gaussian components.  The data hint that
deviations from Gaussianity may diminish for bright ($> 30$~Jy)
masers, but our sample size of bright masers is too small to be
conclusive.  Maser FWHM linewidths range from 0.15 to 0.38~\kms, with
1665 MHz masers generally having broader profiles than other
ground-state masers.

Consistent with theoretical predictions \citep{goldreich73b}, we do
not see $\sigma^{\pm 2,3}$ components in the 1612 and 1720 MHz
satellite-line transitions.  When satellite-line Zeeman pairs are
seen, the magnetic fields are most consistent with values derived from
main-line transitions if the splitting appropriate to $\sigma^{\pm 1}$
components is assumed.

Velocity gradients are common in OH masers.  In W3(OH), 1667 MHz
masers are seen to have large positional gradients (i.e., the position
in the plane of the sky changes rapidly as a function of LSR
velocity), corresponding to small velocity gradients.  This is
consistent with predictions by \citet{pavlakis96}, who find that small
velocity gradients are required for significant amplification at 1667
MHz.

Maser spot sizes appear to be larger in the line wings than at line
center.  The increase of deconvolved spot size with frequency offset
from center is small enough to argue against a spherical maser
geometry \citep{elitzur90}.  However, data of higher sensitivity and
spatial resolution are required to conclusively argue for or against
specific maser geometries.

\acknowledgements

We thank J.~Romney, C.~Walker, \& L.~Foley for their assistance in
identifying allowed modes of operation of the Socorro correlator.

{\it Facility: VLBA}

\clearpage
\LongTables
\begin{deluxetable}{rccrrcrrrr}
\tabletypesize{\tiny}
\tablecaption{Detected Masers in W3(OH)\label{tab-spots}}
\tablehead{
  \colhead{} &
  \colhead{} &
  \colhead{} &
  \colhead{} &
  \colhead{} &
  \colhead{} &
  \colhead{} &
  \colhead{Velocity} &
  \colhead{} &
  \colhead{} \\
  \colhead{Spot} &
  \colhead{Freq.} &
  \colhead{} &
  \colhead{RA Offset\tnm{a}} &
  \colhead{Dec Offset\tnm{a}} &
  \colhead{Velocity\tnm{b}} &
  \colhead{Brightness\tnm{c}}&
  \colhead{Gradient} &
  \colhead{PA\tnm{d}} &
  \colhead{Zeeman} \\
  \colhead{Number} &
  \colhead{(MHz)} &
  \colhead{Pol.} &
  \colhead{(mas)} &
  \colhead{(mas)} &
  \colhead{(\kms)} &
  \colhead{(Jy beam$^{-1}$)} &
  \colhead{(\kms~mas$^{-1}$)} &
  \colhead{(\degr)} &
  \colhead{Pair\tnm{e}}
}
\startdata
  1& 1665 & L & $-$1401.01 &  $-$428.95 &$-$45.71  &   1.35 &\nodata&\nodata&\nodata\\
  2& 1665 & L & $-$1083.78 &  $-$229.84 &$-$46.26  &   2.66 & 0.051 & $-$43 &\nodata\\
  3& 1665 & L & $-$1046.72 &  $-$128.30 &$-$46.00  &   2.22 &\nodata&\nodata&\nodata\\
  4& 1665 & L & $-$1046.26 &     134.24 &$-$45.85  &   1.12 &\nodata&\nodata&\nodata\\
  5& 1665 & R & $-$1044.57 &     133.00 &$-$45.82  &   0.71 & 0.092 &   150 &\nodata\\
  6& 1665 & L & $-$1025.43 &  $-$121.83 &$-$46.22  &  72.32 & 0.052 &$-$106 &\nodata\\
  7& 1665 & L & $-$1017.24 &  $-$121.19 &$-$46.24  &   7.07 & 0.034 &$-$106 &\nodata\\
  8& 1665 & L &  $-$963.38 &   $-$60.68 &$-$45.43  &   6.77 & 0.080 &   137 &\nodata\\
  9& 1665 & L &  $-$960.09 &  $-$101.69 &$-$46.46  &  56.64 & 0.258 &     9 &     1 \\
 10& 1665 & R &  $-$955.75 &   $-$96.08 &$-$43.36  &   1.75 & 0.110 &    10 &     1 \\
 11& 1665 & R &  $-$949.96 &      30.40 &$-$44.53  &   1.53 & 0.051 &   170 &\nodata\\
 12& 1665 & R &  $-$941.76 &      28.11 &$-$44.48  &   0.77 &\nodata&\nodata&\nodata\\
 13& 1665 & L &  $-$936.04 &  $-$146.42 &$-$46.22  &   2.26 &\nodata&\nodata&\nodata\\
 14& 1665 & L &  $-$910.16 &     118.12 &$-$45.21  &   7.24 &\nodata&\nodata&\nodata\\
 15& 1665 & L &  $-$908.42 &     103.93 &$-$45.01  &   8.67 & 0.091 &    49 &\nodata\\
 16& 1665 & L &  $-$905.02 &     121.54 &$-$45.19  &   6.17 & 0.109 &   137 &\nodata\\
 17& 1665 & R &  $-$904.18 &     104.96 &$-$45.01  &  87.25 & 0.039 &$-$137 &\nodata\\
 18& 1665 & R &  $-$903.84 &     106.50 &$-$44.92  &  86.66 & 0.112 &    99 &\nodata\\
 19& 1665 & R &  $-$903.43 &     122.38 &$-$45.16  &  81.34 & 0.086 &$-$115 &\nodata\\
 20& 1665 & L &  $-$903.21 &      43.38 &$-$47.10  &   8.78 & 0.162 & $-$65 &     2 \\
 21& 1665 & L &  $-$898.25 &      69.41 &$-$47.91  &   0.23 &\nodata&\nodata&     3 \\
 22& 1665 & R &  $-$897.89 &      42.91 &$-$43.82  &   0.41 &\nodata&\nodata&     2 \\
 23& 1665 & R &  $-$896.86 &      47.56 &$-$44.13  &   0.53 &\nodata&\nodata&\nodata\\
 24& 1665 & R &  $-$895.04 & $-$1661.62 &$-$42.70  &   0.30 &\nodata&\nodata&\nodata\\
 25& 1665 & R &  $-$894.15 &      72.11 &$-$44.22  &   2.25 & 0.263 &   112 &     3 \\
 26& 1665 & R &  $-$891.01 & $-$1672.87 &$-$42.31  &   0.40 &\nodata&\nodata&\nodata\\
 27& 1665 & R &  $-$889.81 &  $-$591.76 &$-$44.44  &   2.66 & 0.033 & $-$36 &\nodata\\
 28& 1665 & L &  $-$880.24 &      66.78 &$-$47.12  &   1.41 & 0.138 &    42 &\nodata\\
 29& 1665 & R &  $-$864.21 &  $-$621.18 &$-$44.15  &   1.55 & 0.260 &    79\tnm{f} &\nodata\\
 30& 1667 & L &  $-$863.77 &  $-$652.52 &$-$45.58  &   0.79 & 0.028 &$-$173 &     4 \\
 31& 1667 & R &  $-$863.75 &  $-$652.51 &$-$44.70  &   1.68 & 0.038 &   153 &     4 \\
 32& 1665 & L &  $-$861.15 &  $-$641.00 &$-$46.11  &   2.69 &\nodata&\nodata&     5 \\
 33& 1665 & R &  $-$856.54 &  $-$648.96 &$-$44.20  &  17.31 & 0.038 &     6 &     5 \\
 34& 1665 & L &  $-$853.09 & $-$1733.38 &$-$44.90  &   5.22 & 0.171 &    86 &\nodata\\
 35& 1665 & R &  $-$852.27 &  $-$655.96 &$-$44.07  &   2.68 &\nodata&\nodata&\nodata\\
 36& 1665 & L &  $-$851.27 &  $-$617.71 &$-$45.76  &   4.12 & 0.080 & $-$19 &     6 \\
 37& 1665 & R &  $-$848.73 &  $-$617.03 &$-$43.93  &   4.50 & 0.260 &    79\tnm{f} &     6 \\
 38& 1665 & L &  $-$848.47 & $-$1732.56 &$-$44.94  &  12.04 & 0.173 & $-$93 &\nodata\\
 39& 1665 & R &  $-$833.29 &  $-$678.88 &$-$44.09  &   1.22 & 0.022 &  $-$4 &\nodata\\
 40& 1665 & R &  $-$828.87 &  $-$713.52 &$-$44.26  &   2.52 & 0.040 &   173 &\nodata\\
 41& 1665 & L &  $-$755.01 & $-$1693.36 &$-$45.19  &   9.83 & 0.065 & $-$56 &     7 \\
 42& 1667 & L &  $-$739.54 & $-$1809.93 &$-$44.07  &   0.34 &\nodata&\nodata&\nodata\\
 43& 1667 & L &  $-$738.36 & $-$1891.89 &$-$43.74  &   0.60 & 0.079 &   152 &\nodata\\
 44& 1665 & R &  $-$738.26 & $-$1698.87 &$-$42.84  &   2.27 & 0.075 & $-$99 &     7 \\
 45& 1665 & L &  $-$733.72 & $-$1887.42 &$-$44.55  &   0.92 &\nodata&\nodata&\nodata\\
 46& 1665 & R &  $-$730.81 & $-$1678.82 &$-$42.88  &   4.86 & 0.087 &$-$121 &     8 \\
 47& 1665 & L &  $-$725.64 & $-$1679.43 &$-$44.46  &   2.31 & 0.046 &$-$126 &     8 \\
 48& 1665 & R &  $-$720.78 & $-$1675.72 &$-$44.46  &   1.78 & 0.073 &$-$114 &\nodata\\
 49& 1665 & L &  $-$717.11 &  $-$345.82 &$-$46.84  &   0.32 &\nodata&\nodata&     9 \\
 50& 1665 & R &  $-$715.68 &  $-$533.45 &$-$44.07  &   0.37 &\nodata&\nodata&\nodata\\
 51& 1665 & R &  $-$713.96 &  $-$344.44 &$-$44.42  &   1.52 & 0.833 & $-$48 &     9 \\
 52& 1665 & L &  $-$711.23 & $-$1672.73 &$-$45.65  &  17.76 & 0.242 &    97 &\nodata\\
 53& 1665 & R &  $-$710.98 & $-$1671.67 &$-$45.63  &   0.21 &\nodata&\nodata&\nodata\\
 54& 1667 & L &  $-$699.48 & $-$1883.25 &$-$44.66  &   2.74 & 0.065 &   119 &    10 \\
 55& 1665 & R &  $-$699.17 & $-$1626.33 &$-$40.55  &   2.66 & 0.156 &    63 &\nodata\\
 56& 1667 & R &  $-$697.74 & $-$1878.05 &$-$42.31  &   3.33 & 0.044 &   114 &    10 \\
 57& 1665 & R &  $-$691.44 & $-$1873.41 &$-$41.19  &  11.57 & 0.059 &  $-$8 &\nodata\\
 58& 1667 & L &  $-$677.26 & $-$1895.84 &$-$44.44  &   9.92 & 0.063 & $-$57 &    11 \\
 59& 1667 & R &  $-$675.79 & $-$1890.67 &$-$42.16  &  10.65 & 0.104 & $-$31 &    11 \\
 60& 1665 & R &  $-$657.34 & $-$1899.16 &$-$41.41  &   0.63 &\nodata&\nodata&\nodata\\
 61& 1667 & L &  $-$656.52 & $-$1912.13 &$-$44.59  &   5.60 & 0.063 &$-$163 &    12 \\
 62& 1667 & R &  $-$656.38 & $-$1904.87 &$-$42.27  &   2.41 & 0.058 &   168 &    12 \\
 63& 1665 & R &  $-$648.75 & $-$1908.53 &$-$41.45  &   0.72 &\nodata&\nodata&\nodata\\
 64& 1665 & L &  $-$637.46 & $-$1549.28 &$-$45.14  &   1.19 &\nodata&\nodata&\nodata\\
 65& 1667 & L &  $-$626.45 & $-$1919.08 &$-$44.66  &   1.24 &\nodata&\nodata&\nodata\\
 66& 1665 & R &  $-$622.58 & $-$1918.34 &$-$41.36  &   0.53 &\nodata&\nodata&\nodata\\
 67& 1667 & L &  $-$606.23 & $-$1914.63 &$-$44.81  &   5.45 & 0.039 &    82 &\nodata\\
 68& 1665 & L &  $-$598.81 & $-$1921.76 &$-$45.49  &   2.58 &\nodata&\nodata&\nodata\\
 69& 1667 & R &  $-$597.38 & $-$1917.93 &$-$42.49  &   4.28 & 0.038 &   109 &    13 \\
 70& 1667 & L &  $-$596.38 & $-$1922.68 &$-$44.81  &   9.09 & 0.053 &   118 &    13 \\
 71& 1665 & L &  $-$593.99 & $-$1917.48 &$-$45.52  &  15.82 & 0.735 &   150 &    14 \\
 72& 1665 & R &  $-$591.28 & $-$1915.88 &$-$41.58  &  10.11 & 0.345 &    70 &    14 \\
 73& 1667 & L &  $-$571.99 & $-$1935.16 &$-$44.53  &   0.95 &\nodata&\nodata&\nodata\\
 74& 1667 & L &  $-$569.43 & $-$1752.99 &$-$44.66  &   0.72 & 0.037 &    76 &    15 \\
 75& 1667 & R &  $-$560.66 & $-$1750.80 &$-$42.29  &   0.60 &\nodata&\nodata&    15 \\
 76& 1667 & L &  $-$554.80 & $-$1948.10 &$-$44.09  &   0.42 &\nodata&\nodata&\nodata\\
 77& 1667 & L &  $-$549.71 & $-$1764.13 &$-$44.59  &   0.69 & 0.139 &$-$124 &\nodata\\
 78& 1667 & L &  $-$544.63 & $-$1746.85 &$-$45.87  &   1.02 & 0.055 &    84 &\nodata\\
 79& 1665 & L &  $-$525.45 & $-$1778.23 &$-$46.26  &  14.98 & 0.299 &    14 &\nodata\\
 80& 1665 & R &  $-$519.88 & $-$1778.09 &$-$42.68  &   0.77 &\nodata&\nodata&    16 \\
 81& 1665 & L &  $-$519.28 & $-$1773.84 &$-$45.45  &   8.59 & 0.071 & $-$23\tnm{f} &    16 \\
 82& 1665 & L &  $-$517.36 & $-$1781.08 &$-$46.02  &   7.97 & 0.071 & $-$23\tnm{f} &    17 \\
 83& 1667 & L &  $-$515.12 & $-$1905.69 &$-$44.59  &   3.00 & 0.031 &$-$143 &    18 \\
 84& 1667 & R &  $-$514.49 & $-$1897.95 &$-$42.22  &   0.99 & 0.095 &$-$114 &    18 \\
 85& 1665 & R &  $-$513.26 & $-$1781.87 &$-$42.75  &   1.02 &\nodata&\nodata&    17 \\
 86& 1665 & L &  $-$500.05 & $-$1787.14 &$-$44.57  &   7.30 & 0.063 &   106\tnm{f} &\nodata\\
 87& 1665 & L &  $-$493.83 & $-$1788.52 &$-$44.24  &   8.08 & 0.063 &   106\tnm{f} &    19 \\
 88& 1665 & L &  $-$493.18 & $-$1772.50 &$-$45.16  &   5.23 &\nodata&\nodata&\nodata\\
 89& 1665 & R &  $-$489.49 & $-$1787.39 &$-$40.07  &   1.20 & 0.171 &    94 &    19 \\
 90& 1665 & R &  $-$488.77 & $-$1787.03 &$-$44.07  &   3.53 & 0.144 &   108 &\nodata\\
 91& 1667 & L &  $-$487.24 & $-$1802.74 &$-$43.65  &   0.22 &\nodata&\nodata&\nodata\\
 92& 1665 & L &  $-$484.31 & $-$1778.07 &$-$45.38  &  14.73 & 0.197 &     4\tnm{f}&\nodata\\
 93& 1665 & L &  $-$482.02 & $-$1777.17 &$-$45.16  &  12.04 & 0.197 &     4\tnm{f}&    20 \\
 94& 1665 & R &  $-$481.92 & $-$1773.12 &$-$45.14  &   0.60 &\nodata&\nodata&\nodata\\
 95& 1665 & R &  $-$479.01 & $-$1777.68 &$-$41.74  &   0.61 &\nodata&\nodata&    20 \\
 96& 1665 & R &  $-$475.48 & $-$1808.55 &$-$44.59  &   0.52 &\nodata&\nodata&\nodata\\
 97& 1667 & R &  $-$429.02 & $-$1803.49 &$-$44.02  &   1.54 & 0.045 &   136 &\nodata\\
 98& 1667 & L &  $-$395.12 & $-$1895.44 &$-$43.93  &   1.42 &\nodata&\nodata&\nodata\\
 99& 1667 & L &  $-$388.20 & $-$1883.45 &$-$44.28  &  21.32 & 0.071 & $-$99 &\nodata\\
100& 1667 & R &  $-$387.38 & $-$1878.44 &$-$44.26  &   5.39 & 0.091 & $-$95 &\nodata\\
101& 1665 & R &  $-$350.57 & $-$1177.07 &$-$43.30  &   2.01 & 0.083 &$-$149 &\nodata\\
102& 1667 & R &  $-$345.28 & $-$1873.60 &$-$44.02  &   0.47 & 0.088 &     7 &\nodata\\
103& 1667 & R &  $-$342.90 & $-$1874.70 &$-$41.94  &   0.24 &\nodata&\nodata&\nodata\\
104& 1665 & R &  $-$310.81 & $-$1857.16 &$-$44.92  &   3.86 & 0.121 & $-$78 &\nodata\\
105& 1665 & L &  $-$302.88 & $-$1428.48 &$-$47.30  &   0.27 &\nodata&\nodata&    21 \\
106& 1665 & R &  $-$301.89 & $-$1075.94 &$-$42.66  &  10.64 & 0.186 &    40 &\nodata\\
107& 1665 & R &  $-$301.29 & $-$1429.48 &$-$43.93  &   0.90 & 0.063 &    41 &    21 \\
108& 1665 & R &  $-$299.89 &  $-$842.63 &$-$43.43  &   0.47 & 0.046 &   157\tnm{f}&\nodata\\
109& 1665 & R &  $-$299.23 &  $-$847.67 &$-$43.34  &   0.43 & 0.046 &   157\tnm{f}&\nodata\\
110& 1665 & R &  $-$291.66 & $-$1870.98 &$-$45.80  &   0.24 &\nodata&\nodata&\nodata\\
111& 1665 & R &  $-$288.38 &  $-$757.42 &$-$43.30  &   0.39 &\nodata&\nodata&\nodata\\
112& 1665 & R &  $-$287.09 & $-$1011.08 &$-$42.37  &   0.27 &\nodata&\nodata&\nodata\\
113& 1665 & R &  $-$283.89 &  $-$564.15 &$-$42.94  &   0.38 & 0.108 &   148 &\nodata\\
114& 1665 & R &  $-$275.26 &  $-$746.67 &$-$42.70  &   0.98 & 0.065 &    16 &\nodata\\
115& 1665 & L &  $-$272.93 & $-$1783.63 &$-$44.81  &   1.79 & 0.119 &$-$127 &\nodata\\
116& 1665 & R &  $-$256.48 &  $-$690.40 &$-$42.59  &   0.62 & 0.074 &$-$159 &\nodata\\
117& 1665 & R &  $-$245.22 &  $-$703.98 &$-$42.33  &   0.85 & 0.073 &    38 &\nodata\\
118& 1667 & L &  $-$242.05 & $-$1999.62 &$-$43.28  &   0.33 &\nodata&\nodata&\nodata\\
119& 1665 & L &  $-$222.24 &  $-$544.71 &$-$45.98  &  10.10 & 1.266 & $-$42 &\nodata\\
120& 1665 & L &  $-$212.80 &  $-$573.76 &$-$45.65  &   3.78 & 0.149 &$-$157 &    22 \\
121& 1665 & R &  $-$209.67 &  $-$571.73 &$-$41.85  &   0.27 &\nodata&\nodata&    22 \\
122& 1665 & L &  $-$168.26 &   $-$43.45 &$-$46.40  &   8.61 & 0.325 &   135 &\nodata\\
123& 1665 & L &  $-$166.44 & $-$1131.04 &$-$45.80  &   0.85 &\nodata&\nodata&    23 \\
124& 1665 & R &  $-$165.71 &   $-$42.27 &$-$46.35  &   0.27 &\nodata&\nodata&\nodata\\
125& 1665 & R &  $-$163.42 & $-$1128.55 &$-$42.86  &   5.41 & 0.189 &   110 &    23 \\
126& 1720 & L &  $-$160.40 & $-$1113.22 &$-$44.55  &   0.17 &\nodata&\nodata&    24 \\
127& 1720 & R &  $-$160.08 & $-$1114.71 &$-$43.45  &   0.90 &\nodata&\nodata&    24 \\
128& 1665 & L &  $-$154.38 & $-$1777.08 &$-$43.36  &   0.51 & 0.258 &     3 &\nodata\\
129& 1665 & L &  $-$150.95 & $-$1169.08 &$-$44.42  &   2.86 &\nodata&\nodata&\nodata\\
130& 1720 & R &  $-$150.68 & $-$1122.99 &$-$43.17  &   2.62 & 0.046 &   133 &    25 \\
131& 1720 & L &  $-$150.09 & $-$1122.69 &$-$43.77  &   2.40 & 0.068 &   140 &    25 \\
132& 1665 & L &  $-$146.90 & $-$1182.53 &$-$44.59  &   5.79 & 0.202 &$-$105 &\nodata\\
133& 1665 & R &  $-$146.78 &   $-$59.45 &$-$40.29  &   0.40 & 0.167 &   173 &\nodata\\
134& 1665 & L &  $-$141.02 & $-$1785.66 &$-$43.32  &   0.23 &\nodata&\nodata&\nodata\\
135& 1665 & L &  $-$140.86 & $-$1190.35 &$-$44.70  &   7.28 & 0.118 & $-$25 &    26 \\
136& 1665 & R &  $-$138.11 & $-$1177.48 &$-$41.74  &   7.76 & 0.050 &   169\tnm{f}&\nodata\\
137& 1665 & R &  $-$137.79 & $-$1189.18 &$-$41.32  &   3.59 & 0.050 &   169\tnm{f}&    26 \\
138& 1665 & L &  $-$111.41 & $-$1331.05 &$-$45.14  &   2.35 &\nodata&\nodata&    27 \\
139& 1665 & R &  $-$108.21 & $-$1330.27 &$-$41.69  &   0.77 &\nodata&\nodata&    27 \\
140& 1665 & L &  $-$106.91 &   $-$56.41 &$-$45.98  &   3.22 & 0.392 & $-$38 &    28 \\
141& 1665 & R &  $-$103.64 &   $-$56.02 &$-$39.74  &   0.19 &\nodata&\nodata&    28 \\
142& 1665 & L &   $-$98.12 &     252.88 &$-$48.99  &   0.15 &\nodata&\nodata&\nodata\\
143& 1665 & L &   $-$95.56 &   $-$62.56 &$-$46.50  &   3.38 & 0.152 & $-$39 &\nodata\\
144& 1665 & L &   $-$95.43 &      26.74 &$-$47.05  &   2.15 & 0.291 &$-$169 &\nodata\\
145& 1665 & R &   $-$92.17 &      29.49 &$-$47.14  &   0.56 & 0.105 &$-$169 &\nodata\\
146& 1665 & L &   $-$90.08 &   $-$68.05 &$-$46.37  &   6.32 & 2.000 &$-$149 &\nodata\\
147& 1665 & R &   $-$84.13 &  $-$183.84 &$-$41.03  &   0.69 &\nodata&\nodata&\nodata\\
148& 1667 & L &   $-$82.81 & $-$2033.81 &$-$47.60  &   1.50 & 0.024 &$-$131\tnm{f} &\nodata\\
149& 1665 & L &   $-$80.56 & $-$1753.66 &$-$41.96  &   1.57 & 0.199 &$-$118 &\nodata\\
150& 1665 & R &   $-$74.05 & $-$1403.99 &$-$41.82  &   0.53 &\nodata&\nodata&\nodata\\
151& 1667 & R &   $-$72.53 & $-$2014.12 &$-$47.80  &   0.36 &\nodata&\nodata&\nodata\\
152& 1667 & L &   $-$72.03 & $-$2022.71 &$-$47.80  &   1.54 & 0.024 &$-$131\tnm{f}&\nodata\\
153& 1612 & R &   $-$67.38 &  $-$216.13 &$-$42.43  &   1.05 & 0.235 &    67 &    29 \\
154& 1612 & L &   $-$66.79 &  $-$215.93 &$-$43.46  &   0.81 & 10.00 &   173 &    29 \\
155& 1667 & R &   $-$61.65 & $-$1411.27 &$-$42.75  &   1.08 & 0.095 & $-$11 &\nodata\\
156& 1665 & L &   $-$58.58 &     235.18 &$-$49.08  &   0.17 &\nodata&\nodata&\nodata\\
157& 1665 & L &   $-$56.38 & $-$1972.37 &$-$48.28  &   0.26 &\nodata&\nodata&    30 \\
158& 1665 & L &   $-$55.29 &  $-$189.59 &$-$44.59  &   0.64 &\nodata&\nodata&    31 \\
159& 1665 & R &   $-$55.10 & $-$1972.51 &$-$46.88  &   0.34 & 0.044 &   167 &    30 \\
160& 1665 & R &   $-$54.48 &  $-$187.06 &$-$40.42  &   0.74 & 1.667 &    88 &    31 \\
161& 1665 & L &   $-$54.00 &   $-$91.17 &$-$46.84  &   1.91 & 0.513 &$-$148 &\nodata\\
162& 1665 & R &   $-$53.31 & $-$1969.86 &$-$48.24  &   0.32 & 0.058 & $-$97 &\nodata\\
163& 1665 & L &   $-$51.69 &   $-$81.33 &$-$44.99  &   2.41 & 0.446 & $-$40 &\nodata\\
164& 1665 & L &   $-$49.10 &  $-$191.72 &$-$45.30  &   7.37 & 0.398 & $-$44 &    32 \\
165& 1612 & R &   $-$46.71 &  $-$210.61 &$-$43.14  &   0.24 & 0.060 &$-$134 &    34 \\
166& 1665 & R &   $-$46.25 &  $-$189.74 &$-$40.62  &   1.08 & 0.592 &  $-$2 &    32 \\
167& 1665 & L &   $-$46.04 & $-$1748.63 &$-$43.54  &   0.43 &\nodata&\nodata&    33 \\
168& 1667 & R &   $-$45.84 & $-$1431.56 &$-$42.95  &   1.73 & 0.111 &    14 &    36 \\
169& 1612 & L &   $-$45.05 &  $-$210.30 &$-$44.09  &   0.26 &\nodata&\nodata&    34 \\
170& 1665 & R &   $-$41.85 &   $-$50.36 &$-$43.41  &   0.77 & 0.296 &  $-$4 &    35 \\
171& 1665 & L &   $-$41.80 &  $-$192.21 &$-$45.67  &   1.77 &\nodata&\nodata&\nodata\\
172& 1665 & L &   $-$41.32 &   $-$98.62 &$-$48.31  &   0.35 & 0.081 &  $-$8 &    35 \\
173& 1667 & L &   $-$40.06 & $-$1445.19 &$-$45.25  &   0.27 &\nodata&\nodata&    36 \\
174& 1665 & R &   $-$39.71 & $-$1748.07 &$-$40.88  &   0.45 &\nodata&\nodata&    33 \\
175& 1665 & R &   $-$30.61 & $-$1742.49 &$-$41.16  &   4.99 & 0.129 &$-$129 &    37 \\
176& 1665 & L &   $-$29.57 & $-$1744.27 &$-$43.98  &  10.90 & 0.083 &$-$118 &    37 \\
177& 1665 & R &   $-$23.96 & $-$1729.29 &$-$40.90  &   1.52 & 0.193 &$-$103 &    38 \\
178& 1665 & L &   $-$21.49 & $-$1737.53 &$-$44.11  &   5.40 & 0.151 &$-$130 &    38 \\
179& 1665 & R &   $-$17.98 &  $-$119.70 &$-$40.46  &   0.19 &\nodata&\nodata&\nodata\\
180& 1665 & R &   $-$12.31 &  $-$120.95 &$-$39.58  &   0.22 &\nodata&\nodata&\nodata\\
181& 1665 & L &    $-$8.58 &   $-$16.35 &$-$47.87  &   3.13 & 0.309 &    15 &    39 \\
182& 1665 & L &    $-$8.54 &    $-$2.84 &$-$48.53  &   4.24 & 0.238 &   120 &    40 \\
183& 1612 & L &    $-$7.19 &  $-$116.00 &$-$43.02  &   1.28 & 0.442 & $-$39 &    41 \\
184& 1665 & R &    $-$7.06 &   $-$14.74 &$-$41.63  &  33.31 & 0.352 &   123 &    39 \\
185& 1612 & R &    $-$6.27 &  $-$115.21 &$-$41.78  &   0.20 &\nodata&\nodata&    41 \\
186& 1665 & R &    $-$5.16 &    $-$0.07 &$-$48.48  &   1.46 & 0.265 &   113 &\nodata\\
187& 1665 & L &    $-$5.14 &      23.68 &$-$44.50  &   0.92 & 0.296 &    43 &\nodata\\
188& 1665 & R &    $-$5.08 &    $-$1.19 &$-$43.25  &   0.28 &\nodata&\nodata&    40 \\
189& 1665 & R &    $-$5.01 &   $-$19.74 &$-$40.86  &   0.26 &\nodata&\nodata&\nodata\\
190& 1720 & R &    $-$2.82 &      27.60 &$-$44.89  &  11.66 & 0.135 &    46 &    42 \\
191& 1665 & R &    $-$2.01 &      25.34 &$-$44.48  &   1.80 & 1.031 &     4 &\nodata\\
192& 1720 & L &    $-$2.01 &      28.73 &$-$45.64  &  14.63 & 0.111 &    39 &    42 \\
193& 1665 & L &    $-$0.72 &       0.53 &$-$47.47  &  92.03 & 1.333 &   118 &    43 \\
194& 1665 & R &       1.43 & $-$1719.83 &$-$41.17  &   0.89 & 0.097 &    25 &\nodata\\
195& 1665 & R &       1.75 &       6.30 &$-$42.88  &   0.17 &\nodata&\nodata&    43 \\
196& 1665 & R &       2.45 &       2.93 &$-$47.47  &  68.35 & 1.042 &   146 &\nodata\\
197& 1665 & L &       3.89 &   $-$81.13 &$-$44.48  &   0.62 &\nodata&\nodata&    44 \\
198& 1665 & L &       7.06 &  $-$131.47 &$-$45.58  &   6.98 & 0.272 &$-$160 &\nodata\\
199& 1665 & L &       8.02 &  $-$107.91 &$-$45.60  &   4.85 & 0.331 &    64 &\nodata\\
200& 1665 & L &      10.94 &   $-$94.17 &$-$45.63  &   1.60 &\nodata&\nodata&    45 \\
201& 1665 & R &      13.53 &   $-$92.21 &$-$39.10  &   0.14 &\nodata&\nodata&    45 \\
202& 1665 & R &      14.45 &   $-$77.85 &$-$40.18  &   0.20 &\nodata&\nodata&    44 \\
203& 1665 & L &      15.84 &   $-$16.78 &$-$47.65  &   0.38 &\nodata&\nodata&    46 \\
204& 1665 & R &      19.44 &   $-$14.18 &$-$39.17  &   0.38 & 0.174 &   127 &    46 \\
205& 1665 & L &      21.09 &      11.69 &$-$45.56  &   6.25 & 0.649 &    19 &\nodata\\
206& 1665 & L &      22.11 &     396.09 &$-$49.03  &   1.92 & 0.078 &$-$122 &\nodata\\
207& 1665 & R &      23.94 &     398.64 &$-$48.97  &   0.33 & 0.056 & $-$87 &\nodata\\
208& 1720 & R &      26.16 &      50.80 &$-$42.72  &   7.52 & 0.296 &    41 &    47 \\
209& 1720 & L &      26.36 &      51.12 &$-$43.49  &   3.81 & 0.260 &    49 &    47 \\
210& 1665 & L &      26.57 &   $-$46.26 &$-$45.82  &   0.80 &\nodata&\nodata&\nodata\\
211& 1665 & L &      26.64 & $-$1724.87 &$-$42.64  &   1.38 & 0.174 &$-$139 &\nodata\\
212& 1665 & L &      27.15 &      33.98 &$-$44.84  &   2.04 & 0.065 &  $-$9\tnm{f} &    48 \\
213& 1665 & L &      27.45 &   $-$35.59 &$-$45.65  &   1.63 &\nodata&\nodata&    49 \\
214& 1665 & L &      27.60 &      30.66 &$-$45.12  &   2.45 & 0.065 &  $-$9\tnm{f} &\nodata\\
215& 1665 & R &      28.17 & $-$1719.46 &$-$42.40  &   0.74 &\nodata&\nodata&\nodata\\
216& 1665 & L &      28.60 &     394.66 &$-$48.99  &   2.05 & 0.186 &   156 &\nodata\\
217& 1665 & R &      29.39 &   $-$21.02 &$-$41.10  &   0.74 &\nodata&\nodata&\nodata\\
218& 1665 & R &      29.66 &      35.90 &$-$41.06  &  20.40 & 0.156 &   108 &    48 \\
219& 1665 & R &      30.42 &   $-$31.54 &$-$40.73  &   1.35 & 0.089 &   180 &    49 \\
220& 1665 & R &      31.03 &      48.26 &$-$41.43  &   2.01 &\nodata&\nodata&\nodata\\
221& 1665 & R &      38.31 & $-$1724.34 &$-$42.07  &   0.90 & 0.079 & $-$20 &\nodata\\
222& 1665 & R &      41.00 &     379.37 &$-$48.59  &   0.58 & 0.121 &    90 &\nodata\\
223& 1665 & L &      43.67 &     376.91 &$-$48.55  &   0.79 & 0.111 &   102 &    50 \\
224& 1665 & R &      47.24 & $-$1728.08 &$-$41.69  &   3.16 & 0.222 & $-$25 &\nodata\\
225& 1665 & R &      47.62 &     380.95 &$-$45.32  &   0.52 &\nodata&\nodata&    50 \\
226& 1612 & R &      99.30 & $-$1836.03 &$-$42.93  &   1.06 & 0.089 & $-$89 &\nodata\\
227& 1665 & L &     114.18 &  $-$108.92 &$-$45.43  &   0.95 &\nodata&\nodata&\nodata\\
228& 1665 & L &     125.05 & $-$1775.95 &$-$45.56  &   2.54 & 0.031 &   112 &    51 \\
229& 1665 & R &     126.35 & $-$1774.71 &$-$42.62  &   0.34 &\nodata&\nodata&    51 \\
230& 1665 & R &     162.46 &     327.66 &$-$44.72  &   0.72 &\nodata&\nodata&\nodata\\
231& 1612 & R &     164.89 & $-$1821.44 &$-$42.27  &   1.09 & 0.081 & $-$98 &    52 \\
232& 1612 & L &     167.42 & $-$1820.63 &$-$43.39  &   0.24 &\nodata&\nodata&    52 \\
233& 1612 & R &     173.31 & $-$1821.33 &$-$42.46  &   2.90 & 0.162 & $-$15 &    53 \\
234& 1612 & L &     173.32 & $-$1821.23 &$-$43.36  &   0.58 &\nodata&\nodata&    53 \\
235& 1612 & R &     177.30 & $-$1820.60 &$-$42.62  &   1.61 &\nodata&\nodata&\nodata\\
236& 1665 & L &     181.25 & $-$1800.91 &$-$44.02  &   0.76 &\nodata&\nodata&\nodata\\
237& 1665 & L &     188.74 & $-$1796.46 &$-$44.11  &   1.44 & 0.075 &$-$103 &\nodata\\
238& 1612 & R &     193.99 & $-$1817.59 &$-$43.21  &   0.14 &\nodata&\nodata&    54 \\
239& 1612 & L &     195.03 & $-$1814.94 &$-$44.09  &   0.18 &\nodata&\nodata&    54 \\
240& 1665 & L &     200.50 & $-$1794.18 &$-$44.42  &   5.78 & 0.084 &$-$104 &    55 \\
241& 1665 & R &     202.45 & $-$1792.95 &$-$41.60  &   0.97 & 0.034 &$-$130 &    55 \\
242& 1665 & R &     205.16 & $-$1792.06 &$-$44.46  &   0.55 &\nodata&\nodata&\nodata\\
243& 1612 & L &     233.03 & $-$1721.65 &$-$43.02  &   0.85 & 0.188 &    18 &\nodata\\
244& 1612 & L &     248.05 & $-$1714.66 &$-$42.25  &   0.18 &\nodata&\nodata&    56 \\
245& 1612 & R &     249.76 & $-$1713.64 &$-$41.53  &   0.44 & 0.041 &    56 &    56 \\
246& 1665 & L &     350.46 & $-$1682.49 &$-$44.42  &   0.46 &\nodata&\nodata&\nodata\\
247& 1665 & R &     353.04 & $-$1677.72 &$-$44.42  &   0.34 &\nodata&\nodata&\nodata\\
248& 1665 & R &     359.38 & $-$1866.07 &$-$41.98  &   0.35 &\nodata&\nodata&\nodata\\
249& 1612 & L &     628.04 & $-$1635.73 &$-$47.54  &   0.42 & 0.104 & $-$31 &\nodata\\
250& 1665 & L &     699.04 & $-$1521.77 &$-$47.71  &   0.26 &\nodata&\nodata&\nodata\\
251& 1665 & R &     703.12 & $-$1566.29 &$-$47.69  &   0.38 &\nodata&\nodata&\nodata\\
252& 1665 & L &     718.86 & $-$1511.38 &$-$47.23  &   0.58 & 0.097 &    43 &\nodata\\
253& 1665 & R &     722.13 & $-$1508.82 &$-$47.25  &   0.19 &\nodata&\nodata&\nodata\\
\enddata
\tnt{a}{Centroid of Gaussian fit in channel of peak emission.  See \S
  \ref{results} for discussion of relative alignment of maser spots at
  different frequencies and polarizations.}
\tnt{b}{{}LSR velocity of channel of peak emission.  Adjacent channel
  separation is approximately 0.02~\kms.}
\tnt{c}{Peak brightness of Gaussian fit in channel of peak emission.}
\tnt{d}{Position angle east of north in direction of increasing
line-of-sight velocity across the maser spot.}
\tnt{e}{Zeeman pair number as listed in Table \ref{tab-zeeman}.}
\tnt{f}{Gradient computed over multiple peaks.}
\notetoeditor{This may be best as an online-only table.}
\end{deluxetable}

\begin{deluxetable}{rrrrrrrrr}
\tabletypesize{\small}
\tablecaption{Zeeman Pairs in W3(OH)\label{tab-zeeman}}
\tablehead{
  \colhead{} &
  \colhead{} &
  \multicolumn{3}{c}{\hrulefill LCP \hrulefill} &
  \multicolumn{3}{c}{\hrulefill RCP \hrulefill} &
  \colhead{} \\
  \colhead{Pair} &
  \colhead{Freq.} &
  \colhead{RA Offset} &
  \colhead{Dec Offset} &
  \colhead{$v_\mathrm{LSR}$} &
  \colhead{RA Offset} &
  \colhead{Dec Offset} &
  \colhead{$v_\mathrm{LSR}$} &
  \colhead{$B$\tnm{a}} \\
  \colhead{Number} &
  \colhead{(MHz)} &
  \colhead{(mas)} &
  \colhead{(mas)} &
  \colhead{(\kms)} &
  \colhead{(mas)} &
  \colhead{(mas)} &
  \colhead{(\kms)} &
  \colhead{(mG)}
}
\startdata
%
  1 &1665 & $-$960.09& $-$101.69  & $-$46.46& $-$955.75&  $-$96.08  & $-$43.36 &  5.3 \\
  2 &1665 & $-$903.21&     43.38  & $-$47.10& $-$897.89&     42.91  & $-$43.82 &  5.6 \\
  3 &1665 & $-$898.25&     69.41  & $-$47.91& $-$894.15&     72.11  & $-$44.22 &  6.3 \\
  4 &1667 & $-$863.77& $-$652.52  & $-$45.58& $-$863.75& $-$652.51  & $-$44.70 &  2.5 \\
  5 &1665 & $-$861.15& $-$641.00  & $-$46.11& $-$856.54& $-$648.96  & $-$44.20 &  3.2 \\
  6 &1665 & $-$851.27& $-$617.71  & $-$45.76& $-$848.73& $-$617.03  & $-$43.93 &  3.1 \\
  7 &1665 & $-$755.01&$-$1693.36  & $-$45.19& $-$738.26&$-$1698.87  & $-$42.84 &  4.0\tnm{b} \\
  8 &1665 & $-$725.64&$-$1679.43  & $-$44.46& $-$730.81&$-$1678.82  & $-$42.88 &  2.7 \\
  9 &1665 & $-$717.11& $-$345.82  & $-$46.84& $-$713.96& $-$344.44  & $-$44.42 &  4.1 \\
 10 &1667 & $-$699.48&$-$1883.25  & $-$44.66& $-$697.74&$-$1878.05  & $-$42.31 &  6.6 \\
 11 &1667 & $-$677.26&$-$1895.84  & $-$44.44& $-$675.79&$-$1890.67  & $-$42.61 &  5.2 \\
 12 &1667 & $-$656.52&$-$1912.13  & $-$44.59& $-$656.38&$-$1904.87  & $-$42.27 &  6.6 \\
 13 &1667 & $-$596.38&$-$1922.68  & $-$44.81& $-$597.38&$-$1917.93  & $-$42.49 &  6.6 \\
 14 &1665 & $-$593.99&$-$1917.48  & $-$45.52& $-$591.28&$-$1915.88  & $-$41.58 &  6.7 \\
 15 &1667 & $-$569.43&$-$1752.99  & $-$44.66& $-$560.66&$-$1750.80  & $-$42.29 &  6.7 \\
 16 &1665 & $-$519.28&$-$1773.84  & $-$45.45& $-$519.88&$-$1778.09  & $-$42.68 &  4.7 \\
 17 &1665 & $-$517.36&$-$1781.08  & $-$46.02& $-$513.26&$-$1781.87  & $-$42.75 &  5.5 \\
 18 &1667 & $-$515.12&$-$1905.69  & $-$44.59& $-$514.49&$-$1897.95  & $-$42.22 &  6.7 \\
 19 &1665 & $-$493.83&$-$1788.52  & $-$44.24& $-$489.49&$-$1787.39  & $-$40.07 &  7.1 \\
 20 &1665 & $-$482.02&$-$1777.17  & $-$45.16& $-$479.01&$-$1777.68  & $-$41.74 &  5.8 \\
 21 &1665 & $-$302.88&$-$1428.48  & $-$47.30& $-$301.29&$-$1429.48  & $-$43.93 &  5.7 \\
 22 &1665 & $-$212.80& $-$573.76  & $-$45.65& $-$209.67& $-$571.73  & $-$41.85 &  6.4 \\
 23 &1665 & $-$166.44&$-$1131.04  & $-$45.80& $-$163.42&$-$1128.55  & $-$42.86 &  5.0 \\
 24 &1720 & $-$160.40&$-$1113.22  & $-$44.55& $-$160.08&$-$1114.71  & $-$43.45 &  9.6 \\
 25 &1720 & $-$150.09&$-$1122.69  & $-$43.77& $-$150.68&$-$1122.99  & $-$43.17 &  5.3 \\
 26 &1665 & $-$140.86&$-$1190.35  & $-$44.70& $-$137.79&$-$1189.18  & $-$41.32 &  5.7 \\
 27 &1665 & $-$111.41&$-$1331.05  & $-$45.14& $-$108.21&$-$1330.27  & $-$41.69 &  5.8 \\
 28 &1665 & $-$106.91&  $-$56.41  & $-$45.98& $-$103.64&  $-$56.02  & $-$39.74 & 10.6 \\
 29 &1612 &  $-$66.79& $-$215.93  & $-$43.46&  $-$67.38& $-$216.13  & $-$42.43 &  8.4 \\
 30 &1665 &  $-$56.38&$-$1972.37  & $-$48.28&  $-$55.10&$-$1972.51  & $-$46.88 &  2.4 \\
 31 &1665 &  $-$55.29& $-$189.59  & $-$44.59&  $-$54.48& $-$187.06  & $-$40.42 &  7.1 \\
 32 &1665 &  $-$49.10& $-$191.72  & $-$45.30&  $-$46.25& $-$189.74  & $-$40.62 &  7.9 \\
 33 &1665 &  $-$46.04&$-$1748.63  & $-$43.54&  $-$39.71&$-$1748.07  & $-$40.88 &  4.5 \\
 34 &1612 &  $-$45.05& $-$210.30  & $-$44.09&  $-$46.71& $-$210.61  & $-$43.14 &  7.8 \\
 35 &1665 &  $-$41.32&  $-$98.62  & $-$48.31&  $-$41.85&  $-$50.36  & $-$43.41 &  8.3\tnm{b} \\
 36 &1667 &  $-$40.06&$-$1445.19  & $-$45.25&  $-$45.84&$-$1431.56  & $-$42.95 &  6.5 \\
 37 &1665 &  $-$29.57&$-$1744.27  & $-$43.98&  $-$30.61&$-$1742.49  & $-$41.16 &  4.8 \\
 38 &1665 &  $-$21.49&$-$1737.53  & $-$44.11&  $-$23.96&$-$1729.29  & $-$40.90 &  5.4 \\
 39 &1665 &   $-$8.58&  $-$16.35  & $-$47.87&   $-$7.06&  $-$14.74  & $-$41.63 & 10.6 \\
 40 &1665 &   $-$8.54&   $-$2.84  & $-$48.53&   $-$5.08&   $-$1.19  & $-$43.25 &  8.9 \\
 41 &1612 &   $-$7.19& $-$116.00  & $-$43.02&   $-$6.27& $-$115.21  & $-$41.78 & 10.2 \\
 42 &1720 &   $-$2.01&     28.73  & $-$45.64&   $-$2.82&     27.60  & $-$44.89 &  6.6 \\
 43 &1665 &   $-$0.72&      0.53  & $-$47.47&      1.75&      6.30  & $-$42.88 &  7.8 \\
 44 &1665 &      3.89&  $-$81.13  & $-$44.48&     14.45&  $-$77.85  & $-$40.18 &  7.3 \\
 45 &1665 &     10.94&  $-$94.17  & $-$45.63&     13.53&  $-$92.21  & $-$39.10 & 11.1 \\
 46 &1665 &     15.84&  $-$16.78  & $-$47.65&     19.44&  $-$14.18  & $-$39.17 & 14.4 \\
 47 &1720 &     26.36&     51.12  & $-$43.49&     26.16&     50.80  & $-$42.72 &  6.8 \\
 48 &1665 &     27.15&     33.98  & $-$44.84&     29.66&     35.90  & $-$41.06 &  6.4 \\
 49 &1665 &     27.45&  $-$35.59  & $-$45.65&     30.42&  $-$31.54  & $-$40.73 &  8.3 \\
 50 &1665 &     43.67&    376.91  & $-$48.55&     47.62&    380.95  & $-$45.32 &  5.5 \\
 51 &1665 &    125.05&$-$1775.95  & $-$45.56&    126.35&$-$1774.71  & $-$42.62 &  5.0 \\
 52 &1612 &    167.42&$-$1820.63  & $-$43.39&    164.89&$-$1821.44  & $-$42.27 &  9.2 \\
 53 &1612 &    173.32&$-$1821.23  & $-$43.36&    173.31&$-$1821.33  & $-$42.46 &  7.4 \\
 54 &1612 &    195.03&$-$1814.94  & $-$44.09&    193.99&$-$1817.59  & $-$43.21 &  7.2 \\
 55 &1665 &    200.50&$-$1794.18  & $-$44.42&    202.45&$-$1792.95  & $-$41.60 &  4.8 \\
 56 &1612 &    248.05&$-$1714.66  & $-$42.25&    249.76&$-$1713.64  & $-$41.53 &  5.9 \\
\enddata
\tnt{a}{Assumes splitting appropriate for $\sigma^{\pm1}$ components
   for 1612 and 1720 MHz transitions.  Positive values indicate
   magnetic fields oriented in the hemisphere pointing away from the
   observer.}
 \tnt{b}{{}Large separation between LCP and RCP components.  These may
   be Zeeman ``cousins'' as defined in \citet{fish06}.}
 \notetoeditor{Would this table be useful in an electronic form as well?}
 \end{deluxetable}

\end{document}